\documentclass[journal,twoside]{IEEEtran}
\usepackage{atbegshi,picture,ifpdf,setspace}
\usepackage[utf8]{inputenc}
\usepackage[T1]{fontenc}
\usepackage[pdftex]{graphicx}
\usepackage[caption=false,font=footnotesize]{subfig}
\usepackage{array}
\usepackage{braket}
\usepackage{units}
\usepackage{siunitx}
\usepackage[cmex10]{amsmath}
\usepackage{bm}
\usepackage{cite}
\usepackage{booktabs}
\usepackage{accents}
\usepackage[slantedGreek,cmintegrals]{newtxmath}
\usepackage[fleqn]{empheq}
\usepackage{tcolorbox}
\usepackage[normalem]{ulem}
\DeclareMathAlphabet{\mathbfsf}{\encodingdefault}{\sfdefault}{bx}{it}
\newcommand{\jj}{\mathrm{j}}
\newcommand{\ee}{\mathrm{e}}
\renewcommand*{\vec}{\bm}

\newcommand{\fun}[2]{#1\!\left(#2\right)}
\newcommand{\herm}{^{\mathrm{H}}}
\newcommand{\abs}[1]{\left\lvert#1\right\rvert}
\newcommand{\norm}[1]{\left\lVert#1\right\rVert}
\usepackage{listings}
\usepackage[OMLmathsfit]{isomath}
\DeclareMathAlphabet{\mathbfsf}{\encodingdefault}{\sfdefault}{bx}{n}
\usepackage{bm}
\newcommand* {\matr} [1]{\mathsfbfit{#1}}
\newcommand{\mvec}[1]{\mathsfbfit{#1}}
\usepackage[acronym,shortcuts, nonumberlist]{glossaries}
\newacronym{aut}{AUT}{antenna under test}
\newacronym{ff}{FF}{far field}
\newacronym{nf}{NF}{near field}
\newacronym{nffft}{NFFFT}{near-field far-field transformation}
\newacronym{fiafta}{FIAFTA}{Fast Irregular Antenna Field Transformation Algorithm}
\newacronym{fft}{FFT}{fast Fourier transform}
\newacronym{if}{IF}{intermediate frequency}
\newacronym{snr}{SNR}{signal to noise ratio}
\newacronym{nwa}{VNA}{vector network analyzer}
\newacronym{dof}{DoFs}{degrees of freedom}
\newacronym{svd}{SVD}{singular value decomposition}
\newacronym{rwg}{RWG}{Rao-Wilton-Glisson}
\newacronym{lo}{LO}{local oscillator}
\usepackage{csquotes}

\makeatletter
\def\ps@IEEEtitlepagestyle{%
  \def\@oddfoot{\mycopyrightnotice}%
  \def\@oddhead{\hbox{}\@IEEEheaderstyle\leftmark\hfil\thepage}\relax
  \def\@evenhead{\@IEEEheaderstyle\thepage\hfil\leftmark\hbox{}}\relax
  \def\@evenfoot{}%
}
\def\mycopyrightnotice{%
  \begin{minipage}{\textwidth}
  \centering \scriptsize
  Copyright~\copyright~2020 IEEE. Personal use of this material is permitted. Permission from IEEE must be obtained for all other uses, in any current or future media, including\\reprinting/republishing this material for advertising or promotional purposes, creating new collective works, for resale or redistribution to servers or lists, or reuse of any copyrighted component of this work in other works by sending a request to pubs-permissions@ieee.org.
  \end{minipage}
}
\makeatother

\AtBeginShipout{\AtBeginShipoutUpperLeft{%
  \put(\dimexpr\paperwidth-1cm\relax,-.55cm){\makebox[0pt][r]{
  \begin{minipage}{\textwidth}
  \centering \scriptsize 
  This is the author's version of an article that has been published in this journal. 
  Changes were made to this version by the publisher prior to publication. 
  \\[0.3ex]
  The final version of record is available at\qquad http://dx.doi.org/10.1109/TAP.2020.3008648
  \end{minipage}
  }}}%
}

\begin{document}

\title{Multi-Frequency Phase Retrieval \\for Antenna Measurements}

\author{Josef~Knapp, \textit{Student Member, IEEE},
        Alexander~Paulus, \textit{Student Member, IEEE},
        Jonas Kornprobst,\\ \textit{Student Member, IEEE},
        Uwe Siart, \textit{Member, IEEE},
        and~Thomas~F.~Eibert, \textit{Senior Member, IEEE}
            \thanks{Manuscript received December 27, 2019; revised May 25, 2020; accepted June 13, 2020; date of this version June 18, 2020. This work was supported by the German Federal Ministry for Economic Affairs and Energy under Grant 50RK1923. \emph{(Corresponding author: Josef Knapp.)}%
            	}%
%            \thanks{}
            \thanks{The authors are with the {Chair of High-Frequency Engineering,} Department of Electrical and Computer Engineering, Technical University of Munich, 80290 Munich, Germany (e-mail: josef.knapp@tum.de, hft.ei@tum.de).}%
            \thanks{Color versions of one or more of the figures in this letter are available online at http://ieeexplore.ieee.org.}%
            \thanks{Digital Object Identifier 10.1109/TAP.2020.3008648}% 
        }% <-this % stops a space

\markboth{IEEE TRANSACTIONS ON ANTENNAS AND PROPAGATION}%
{Knapp \MakeLowercase{\textit{et al.}}: Multi-Frequency Phase Retrieval for Antenna Measurements}

\maketitle

\begin{abstract}

Phase retrieval problems in antenna measurements arise when a reference phase cannot be provided to all measurement locations.
Phase retrieval algorithms require sufficiently many independent measurement samples of the radiated fields to be successful.
Larger amounts of independent data may improve the reconstruction of the phase information from magnitude-only measurements.
We show how the knowledge of relative phases among the spectral components of a modulated signal  
at the individual measurement locations may be employed to reconstruct the relative phases between different measurement locations at all frequencies.
Projection matrices map the estimated phases onto the space of fields possibly generated by equivalent \ac{aut} sources at all frequencies.
In this way, the phase of the reconstructed solution is not only restricted by the measurement samples at one frequency, but by the samples at all frequencies simultaneously.
The proposed method can increase the amount of independent phase information even if all probes are located in the \ac{ff} of the \ac{aut}.
  
\end{abstract} 


\IEEEpeerreviewmaketitle

\section{Introduction}
\glsresetall

\IEEEPARstart{T}{ime-harmonic} \ac{nf} antenna pattern measurements classically  comprise magnitude and phase measurements of the fields radiated by the \ac{aut}~\cite{ludwig_nearfield_1971,hansen_spherical_1988,yaghjian_overview_1986}, where the phase information is important for the calculation of desired \ac{ff} quantities (gain, radiation pattern, etc.) from the measured data. 
The phase measurement requires synchronized transmit and receive signals of probe and \ac{aut}, e.g., by a reference phase signal.
Providing a stable phase reference to all different probe locations may call for elaborate measurement setups at high frequencies or for large \acp{aut}~\cite{geise2019crane}, where 
unavoidable movement of the reference cable may introduce severe phase errors. 
Certain measurement setups\,---\,e.g., with a probe fixed to an unmanned aerial vehicle\,---\,may cause severe complications if the phase reference has to be provided at all probe locations.

In order to avoid intricacies associated with phase measurements, and to render some measurement setups feasible in the first place, it is desirable to retrieve the phase information from magnitude-only field measurements, which is a non-linear and non-convex problem in general\,---\,and hard to solve~\cite{isernia_phaseless_1994a,zhang_multifrequency_2009,bandeira_saving_2014,davenport_overview_2016,isernia_phase_1995,sun_geometric_2018,shechtman_phase_2015,burge_phase_1976}.
Since it is attractive in a variety of applications, a large number of algorithms and methods have been proposed to tackle this problem of phase retrieval in magnitude-only antenna measurements~\cite{isernia_phase_1995, isernia_phaseless_1994a,nguyen_reconstruction_1992,pierri_two_1999,yaccarino_phaseless_1999,costanzo_novel_2005,paulus_phaseless_2017,knapp_reconstruction_2019,  bucci_farfield_1990,schmidt_phaseless_2009} or other cases~\cite{kueng_low_2017,candes_phaselift_2013,elser_phase_2003,wu_xray_2005,balan_signal_2006,pfeiffer_phase_2006,bauschke_phase_2002,burge_phase_1976,gerchberg_practical_1972,fienup_phase_1982,nakajima_phase_1987,millane_phase_1990,shechtman_phase_2015,sun_geometric_2018,davenport_overview_2016,bandeira_saving_2014,nakajima_study_1982,alexeev_phase_2014,candes_phase_2015}.
 
In any case, it is understood that one needs to increase the number of measurement samples to enable phase retrieval as compared to the number of samples for conventional measurements with magnitude and phase~\cite{candes_phase_2015,bodmann_stable_2015,huang_phase_2016,waldspurger_phase_2015,eldar_phase_2014,pohl_phaseless_2014,bandeira_saving_2014,isernia_phase_1995, paulus_phaseless_2017, knapp_reconstruction_2019,isernia_phaseless_1994}.
Phase retrieval algorithms usually rely on measurement data with sufficiently many independent measurement samples to allow for a stable reconstruction process. 
With an increasing number of independent measurements, the problem becomes more similar to a convex problem such that even local minimization techniques can be used to find the true solution~\cite{isernia_local_1996, isernia_phaseless_1994,isernia_phaseless_1994a,knapp_comparison_2017}. 
Once the number of independent measurement samples reaches the square of the effective number of unknowns, the problem even can be formulated in a linear manner~\cite{isernia_phase_1989,knapp_reconstruction_2019, candes_phaselift_2013, kueng_low_2017}.
A still unsolved problem is, however, to reliably find a measurement setup which provides sufficiently many independent measurement samples~\cite{maisto_number_2018}. 

Classical attempts employ measurements on two or more surfaces with various distances to the \ac{aut}~\cite{gerchberg_practical_1972,isernia_radiation_1996,fienup_phase_1982, bauschke_phase_2002, bucci_farfield_1990}.
The idea is that the field contributions of the \ac{aut} interfere differently at different distances as long as they are in the \ac{nf} of the \ac{aut}.
The magnitudes convey information about these coherent interferences. 
Since the interference patterns are strongly affected by the phases of individual field contributions, the reconstruction of the relative phases might be possible.
It is still hard to predict where the \ac{nf} measurement locations have to be chosen in order to produce sufficient dissimilar information about the field interferences.  

Specially designed multi-antenna probes have been studied to obtain the required data~\cite{pierri_two_1999,paulus_phaseless_2017,costanzo_integrated_2001}. 
The different probes perform different coherent linear combinations of the incident fields to form their output signals. 
Thus, the magnitudes of the different probe signals heavily depend on the relative phases of incident plane waves and encode the phase information of the incident fields.
The success of this method strongly depends on the possibility of utilizing probes which measure different interference patterns of the radiated \ac{aut} field contributions. 
To be effective, the probes have to be either very large or rather close to the \ac{aut} in order to be able to produce a useful weighted mean of field samples. 
For instance in the \ac{ff}, the field is practically constant over the profile of the probe and taking different linear combinations of this field does not improve the situation.
The amount of different information which can be obtained in this way is also limited by the number of different probes which can be used in the measurement setup.
In the \ac{ff}, the task of providing large numbers of independent measurement samples becomes even more difficult.

In this work, we increase the measurement diversity by introducing relationships between the measured signals at different frequencies. 
No absolute phase reference is required to obtain the relative phases between the measurement samples at different frequencies with the same probe position. 
Phase stability is only required over a short time span in which the measurement samples are obtained for all the frequencies at the individual measurement locations. 
Using the same \ac{lo} for all frequencies at the receiver side (which is not required to be synchronized with the transmit \ac{lo}), it is possible to assign magnitudes and phases to all signal samples at the different frequencies.
The remaining inverse problem is to find the phase differences between the measurement locations (i.e., the phase of the receive \ac{lo} at each position).

Phase retrieval for radiated or scattered fields at multiple frequencies was already investigated occasionally~\cite{costanzo_wideband_2008,arboleya_phaseless_2016,zhang_multifrequency_2009} but these approaches retrieved the phases independently at each frequency or utilized the solution for one frequency as an initial guess at other frequencies.
In contrast to this, we employ the measurement samples at all frequencies simultaneously to find the global phase solution.
Preliminary investigations about this idea can be found in~\cite{paulus_2020_eucap20}.
The source reconstruction problem at one selected reference frequency or a mixture of several reference frequencies is constrained by measured signals at all frequencies.
This constraint is implemented by projection matrices which are designed to remove unphysical portions of the measured signals.
The measurements at all frequencies possibly provide additional information about the relative phases of the radiated fields at the reference frequency.

In Section~\ref{sec:linprob}, the forward problem in antenna measurements is recapitulated for the classical and the magnitude-only problems. 
A possible implementation of a measurement setup which is able to find the phase differences between the signals at different frequencies is described in Section~\ref{sec:setup} in the form of a feasibility study.
In Section~\ref{sec:multifreq}, we show how the measurements at different frequencies can be combined using projection matrices. 
These are used to check whether a given phase solution for the reference frequency is also physically reasonable at the other frequencies.
Section~\ref{sec:step} discusses a rule of thumb for a frequency sampling step size leading to independent data.
In Section~\ref{sec:results}, the performance of the algorithm is investigated for synthetic and measured data.

\section{Notation and Problem Description} \label{sec:linprob}
First, we establish the linear problem to clarify the notation.
For every angular frequency $\omega_k$, an equivalent current method modeling the \ac{aut} by discretized sources~\cite{quijano_field_2010, eibert_multilevel_2009} (e.g., a collection of elementary dipoles) is employed.
The unknown source coefficients are contained in the vector $\mvec{x}_k\in \mathbb{C}^{N \times 1}$.
With increasing frequency, usually more unknowns are utilized to represent the \ac{aut} accurately.
 
The measurement vector corresponding to the frequency $\omega_k$ is denoted by $\mvec{b}_k$ and its $\ell$th entry $[\mvec{b}_k]_\ell$ denotes the measurement sample at the $\ell$th measurement location.
One can establish a linear relationship between the sources and the measurement samples giving rise to the linear equation system~\cite{kornprobst2019solution}
\begin{align}
\mvec{b}_k= \matr{A}_k\mvec{x}_k\, , \label{eq:lin}
\end{align}
with the system matrix $\matr{A}_k \in \mathbb{C}^{M \times N}$.

In a magnitude-only measurement, the phase vector $\mvec{\phi}_k$ with entries defined by
\begin{align}
\ee^{\, \jj [\mvec{\phi}_k]_\ell}\, = \dfrac{[\mvec{b}_k]_\ell}{\abs{[\mvec{b}_k]_\ell}} 
\end{align}
is unknown for every frequency $\omega_k$ and must be determined in a non-linear inverse problem. 
With the elementwise absolute value operator $\abs{\cdot}$, the phase retrieval problem at frequency $\omega_k$ can be expressed as
\begin{align}
\text{find $\mvec{x}_k \in \mathbb{C}^{N \times 1}$ such that} \abs{\mvec{b}_k} = \abs{\matr{A}_k\mvec{x}_k}\label{eq:phaseless_formulation}\,.
\end{align}

In previous works, this phase retrieval problem has been tackled independently for each frequency.
The absolute phases had to be found for every frequency component of the measured signals at all measurement positions separately.
In contrast to this, we assume that the phase differences between signal components at different frequencies are known at each measurement position.
This corresponds to knowing the difference  $\mvec{\phi}_k - \mvec{\phi}_i$ of the phase vectors at frequencies $\omega_k$ and $\omega_i$, respectively.
This assumption simplifies the problem as now only one phase value has to be determined at every measurement position. This phase value determines the absolute phases related to all frequencies within the measured signal at this location.

In the following, we show how this relative phase information can be obtained and exploited for the solution of the phase retrieval problem.

\section{Measurement of Relative Phases of Signals with Different Frequencies}\label{sec:setup}

The following discussions are intended as a proof of concept to demonstrate that multi-frequency antenna measurements are realistic and feasible. 
The aim of this paper is to provide a valid algorithm for the corresponding phase-retrieval problem. Efficient hardware implementations of measurement setups are a topic left for future work.

For most antenna measurement scenarios, phase stability can be ensured up to several tens of \si{GHz} with common technology.
At even higher frequencies, providing a stable phase reference to all measurement locations or \ac{aut} positions becomes challenging and costly.
This section investigates how phase synchronization at a low frequency can suffice to measure relative phases of signals at different frequencies consistently for all measurement locations. 

\begin{figure}
	\centering
	\includegraphics{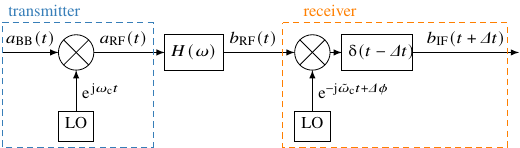}
	\caption{Signal path in a completely asynchronous receiver setup. The receiver \ac{lo} has a frequency shift $\Delta \omega$ and a phase shift $\Delta \phi$ compared to the transmitter \ac{lo}.}\label{fig:system}
\end{figure}     

Consider the case of a totally asynchronous receiver first for which the complete signal path is given in Fig.~\ref{fig:system}\,---later we will use the low frequency base band signal for synchronization.
A certain band limited and periodic input signal
\begin{equation}
\fun{a_{\mathrm{BB}}}{t}=\sum \limits_{k=1}^{N_\mathrm{f}}\alpha_k \ee^{\, \jj \omega_k t}
\end{equation}
with periodicity $T$ consisting of a comb spectrum with a maximum frequency of $\omega_{N\mathrm{f}}$ and spectral coefficients $\alpha_k\in \mathbb{C}$ is upconverted to the carrier frequency $\omega_\mathrm{c}$.
The resulting radio frequency signal 
\begin{equation}
\fun{a_{\mathrm{RF}}}{t}=\sum \limits_{k=1}^{N_\mathrm{f}}\alpha_k \ee^{\, \jj\left(\omega_k+ \omega_\mathrm{c}\right) t}
\end{equation}
is sent from the \ac{aut} to the probe antenna over the measurement channel with transfer function $H(\omega)$.
The receive signal
\begin{equation}
\fun{b_{\mathrm{RF}}}{t}=\sum \limits_{k=1}^{N_\mathrm{f}}\alpha_k \fun{H}{\omega_k+\omega_{\mathrm{c}}} \ee^{\, \jj\left(\omega_k+ \omega_\mathrm{c}\right) t}
\end{equation}
is first downconverted with an asynchronous receiver \ac{lo} (frequency $\tilde{\omega}_\mathrm{c}$) to
\begin{align}
\fun{b_{\mathrm{IF}}}{t}
&=\sum \limits_{k=1}^{N_\mathrm{f}}\alpha_k \fun{H}{\omega_k+\omega_{\mathrm{c}}} \ee^{\, \jj\left(\omega_k+ \omega_\mathrm{c} - \tilde{\omega}_\mathrm{c}\right) t} \ee^{\, \jj \Delta \phi} \notag\\
&=\sum \limits_{k=1}^{N_\mathrm{f}}\alpha_k \fun{H}{\omega_k+\omega_{\mathrm{c}}} \ee^{\, \jj\left(\omega_k+ \Delta \omega\right) t} \ee^{\, \jj \Delta \phi}
\end{align}
and then sampled in time domain.

Compared to the transmitter \ac{lo}, the receiver \ac{lo}  may run at a slightly different frequency $\tilde{\omega}_\mathrm{c}=\omega_\mathrm{c}-\Delta \omega$ and have an unknown phase shift $\Delta \phi$.
Also, in an asynchronous setup, the exact value for $t=0$ is not known at the receiver side. 
Fixing $t=0$ (i.e., the starting time for the sampling interval) to an arbitrary value results in an unknown time shift of $\Delta t$.
The \ac{if} signal is thus written as
\begin{equation}
\fun{b_\mathrm{IF}}{t +\Delta t}= \sum \limits_{k=1}^{N_\mathrm{f}}\alpha_k \fun{H}{\omega_k+\omega_\mathrm{c}} \ee^{\, \jj \left(\omega_k+\Delta\omega\right) t}\ee^{\, \jj \Delta \phi}\ee^{\, \jj \left(\omega_k+\Delta\omega\right) \Delta t}
\end{equation}
and contains all these uncertainties.
Note that the highest frequency which occurs in the signal $b_\mathrm{IF}$ is $\omega_{N_\mathrm{f}}+ \Delta \omega$. 
If the frequencies of the \acp{lo} for up- and downconversion are about the same, $\Delta \omega$ is considerably smaller than $\omega_\mathrm{c}$. 
In order to get rid of the undesired time shift $\Delta t$, it suffices to synchronize the sampling-interval for $b_\mathrm{IF}$ accurately enough for the frequency $\omega_{N_\mathrm{f}}+ \Delta \omega$. 
The periodic base band input signal $a_{\mathrm{BB}}$ can be used for synchronization. 
No technically challenging synchronization of receiver and transmitter around the carrier frequency $\omega_\mathrm{c}$ is required.

\begin{figure}[t]
	\centering
	\includegraphics{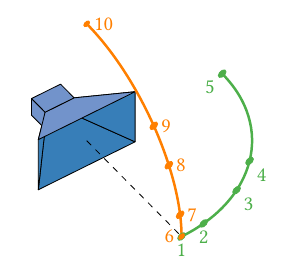}
	\caption{Schematic of the measurement positions for the phase difference measurement verification.} \label{fig:measpos_schematic}
\end{figure}

In order to verify the procedure, measurements have been obtained with an antenna configuration similar as the one in Section~IV-\textit{C}. 
Different channel functions $H(\omega)$ were realized by moving the probe antenna to ten different positions around the \ac{aut} as depicted in Fig.~\ref{fig:measpos_schematic}.

In the considered measurement setup, the in-phase and quadrature components of the base band signal $a_{\mathrm{BB}}$ consisting of 21 spectral components with frequencies from $\SI{-9}{MHz}$ to $\SI{11}{MHz}$ are generated and upconverted ($f_\mathrm{c}= \SI{2.489}{GHz}$) internally with a Rohde \& Schwarz SMBV 100A signal generator~\cite{rohdeschwarz_a}. 
The radio frequency signal $a_{\mathrm{RF}}$ in the range from $\SI{2.480}{GHz}$ to $\SI{2.500}{GHz}$ is sent to the transmitting antenna and downconverted with an external Mini-Circuits 4300 mixer~\cite{minicircuits} after reception by the receive antenna.
The \ac{lo} signal for downconversion ($\tilde{f}_\mathrm{c}= \SI{2.380}{GHz}$) is generated by a separate Rohde \& Schwarz SMR 40 signal generator~\cite{rohdeschwarz_b}. 
No synchronization signal is given to this device, i.e., it constitutes a freely running \ac{lo}.
Finally, the downconverted signal $b_\mathrm{IF}$ together with the in-phase and quadrature component of the base band input signal $a_{\mathrm{BB}}$ are fed to a LeCroy WaveMaster 808Zi-A oscilloscope~\cite{lecroy}. 
For each antenna position, single-shot measurements of the three signals from different times (called “samples” in the remainder of this section) were taken with the oscilloscope.

The magnitude of the base band transmit signal $a_{\mathrm{BB}}$\,---\,which is reconstructed from the measured in-phase and quadrature components\,---\,is used for the synchronization of the measured signals in the post-processing.

\begin{figure}[t]
	\centering
	\includegraphics[width=0.49\textwidth]{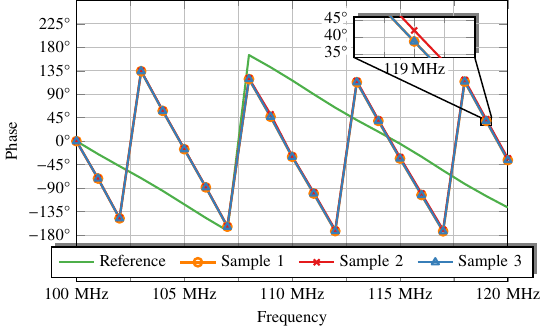}
	\caption{Reconstructed phases dependent on frequency in the IF band at position 1,  relative to the signal phase for the first frequency.} \label{fig:relphase_rel}
\end{figure}

The procedure was carried out for each antenna at  three different times (one sample was taken at a different day than the others) after all devices have been switched off and on again to ensure that no long-term drifting problems occur.
After synchronization, the spectral components of all signals were obtained with a \ac{fft}. The absolute phases of all reconstructed spectral components for the three different measurements differ by a phase $\Delta \phi$ corresponding to the random phase of the receiver \ac{lo}. 
However, the relative phases with respect to the reference frequency at $\SI{100}{MHz}$ show a good agreement. Figure~\ref{fig:relphase_rel} shows the reconstructed relative phases for an exemplary measurement at antenna position 1. 
The overall agreement between the measurements at different times is better than $5^\circ$. The deviations between the oscilloscope measurements and the reference measurements from a vector signal generator (of course taken at the transmit frequencies $\omega_\mathrm{c}+\omega_k$) reveal a systematic deviation due to additional cables which were added to connect the mixer and so on. 

This systematic deviation does not matter for our phase reconstruction algorithm, as a frequency dependent phase offset which is consistently added to all measurement samples does not hurt, since it cancels out during the reconstruction. 
Important is only that the frequency dependent phase bias is the same for all measurement samples such that the phases for all frequencies can be reconstructed up to a global phase bias, if the phase is known at one of the frequencies. 

\begin{figure}[t]
	\centering
	\includegraphics{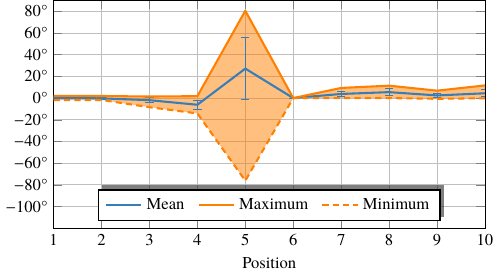}
	\caption{Phase errors of the relative phases at each measurement location.} \label{fig:errors}
\end{figure}

In order to show that the phase bias is constant for all measurement positions, Fig.~\ref{fig:errors} shows the phase errors for every position, after the phase bias from reference position 6 was subtracted. 
The blue line denotes the mean of the phase error over all frequencies and the error bar shows the measured standard deviation. The maximum/minimum values of the phase offset are depicted by the solid and dashed orange lines, respectively. 
For all measurement positions, except position 5, the standard deviation of the reconstructed phase is well below $5^\circ$. 
At position 5, the channel function $H(\omega)$ is at a minimum as it is located at a null of the \ac{aut} pattern. 

\begin{figure}[t]
	\centering
	\includegraphics{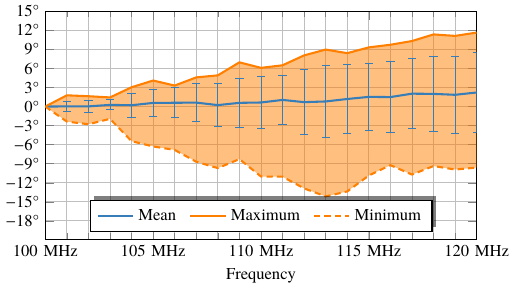}
	\caption{Phase errors for the relative phases of the measurement frequencies in the IF-band.} \label{fig:errors_freq}
\end{figure}

The systematic offset for every measurement frequency (the noisy samples from position 5 were removed as they would obscure the data) is shown in Fig.~\ref{fig:errors_freq}. The continuous increase of the mean deviation suggests that the synchronization can be further improved.  

In conclusion, it was shown that the relative phase difference $\phi_k-\phi_i$ of the signal samples at two frequencies $\omega_k$ and $\omega_i$ of the transfer function between two antennas at high frequencies can be measured in principle for each measurement location without providing a phase reference at the radio frequency.

Important parameters and requirements for the implementation of a reliable system to measure the relative phases between spectral components of a signal are:
\begin{itemize}
	\item the bandwidth of the base band signal $\fun{a_{\mathrm{BB}}}{t}$. 
	Larger bandwidths are preferred for the phase retrieval algorithm described in the sections below.
	However, the phase error between the spectral components due to synchronization errors $\Delta t$ depends linearly on the signal bandwidth, i.e., $\Delta \phi \propto \Delta \omega\, \Delta t$. 
	\item the shape of the base band signal $\fun{a_{\mathrm{BB}}}{t}$. In principle, any periodic  signal with a certain bandwidth can be used, but it helps to have characteristic features in the time domain representation of the signal (such as a steep slope) as well as in the frequency domain representation of the signal (such as discrete frequencies). Such features are beneficial for diminishing synchronization errors $\Delta t$ and frequency offsets of the downconverting mixer at the receiver.
	\item the periodicity $T$ of the base band signal $\fun{a_{\mathrm{BB}}}{t}$. The same signal must be measured at different measurement locations. Averaging over several periods reduces the impact of noise. In a frequency comb spectrum, the frequency step between adjacent frequencies determines the periodicity of the signal.
	\item the \ac{if} frequency range of the downconverted receive signal $\fun{b_\mathrm{IF}}{t}$ and the corresponding sampling frequency should be chosen to support the evaluation of the receive signal with low noise and interference as well as with good accuracy.   
\end{itemize}


\section{The Multi-Frequency Phase Retrieval Problem}\label{sec:multifreq}
The problem to solve is to find the unknown phase terms $\Delta \phi$ for each measurement position, such that a physically reasonable measurement vector $\mvec{b}_k$ is attained for every frequency $\omega_k$, i.e., a measurement vector which could be generated by the given set of \ac{aut} sources.

At a given frequency $\omega_k$, one can perform a projection of any (unphysical) right-hand side vector $\mvec{b}_{k}$ into the space of physically reasonable measurement vectors by multiplying it with the projection matrix $\matr{P}_k=\matr{A}_k^{ }\matr{A}_k^\dagger$ generated with the help of a pseudo-inverse $\matr{A}_k^\dagger$.
To be physically reasonable, the vector $\mvec{b}_k$ must be insensitive to the projection matrix $\matr{P}_k$, i.e., $\mvec{b}_k=\matr{P}_k \mvec{b}_k$.
A physically reasonable vector is orthogonal to the null-space of the projection matrix.
If we assign a false phase to the measured magnitudes, with a high chance, there will be some portion of the reconstructed vector $\mvec{b}_k$ which lies in the null-space of the projection matrix and is, thus, removed.
The projected vector $\tilde{\mvec{b}_k}=\matr{P}_k \mvec{b}_k$ is the closest vector to $\mvec{b}_k$ in the subspace generated by the discretized \ac{aut} sources in $\mvec{x}_k$. 
The dimension of the projection matrix $\matr{P}_k \in \mathbb{C}^{M\times M}$ is independent from the number of unknowns $N$ in the \ac{aut} model and depends only on the number of measurement samples $M$.
Of course, the projection matrices and in particular the required pseudo-matrices are not necessarily constructed explicitly. Instead one may use fast iterative methods to approximate the matrix vector products involving the pseudo-inverses or the corresponding projection matrices.

In the following, a phase retrieval formulation is presented first, which is numerically unstable if measurement values with small magnitudes occur. 
This equation allows to present the main idea in its simplest form.
Subsequently, we will show measures to eradicate the instabilities.
\subsection{Basic Formulation}
With the knowledge of the measured magnitudes and the relative phases, the measurement vector at frequency $\omega_k$ can be expressed by the measurement vector at frequency $\omega_i$ by
\begin{align}
\mvec{b}_k = \matr{U}_{k,i} \mvec{b}_i\, , \label{eq:diag}
\end{align}
where $\matr{U}_{k,i}$ is a diagonal matrix with elements 
\begin{align}
\left[\matr{U}_{k,i}\right]_{\ell\ell}=  \dfrac{\left[\mvec{b}_k\right]_\ell}{\left[\mvec{b}_i\right]_\ell} = \dfrac{\abs{\left[\mvec{b}_k\right]_\ell}}{\abs{\left[\mvec{b}_i\right]_\ell}} \, \ee^{\, \jj \left(\left[\mvec{\phi}_k\right]_\ell-\left[\mvec{\phi}_i\right]_\ell\right)}\, .\label{eq:defU} 
\end{align}
The phase retrieval problems at different frequencies are connected to each other and can be solved simultaneously. 
A certain solution $\mvec{b}_i$ at one frequency $\omega_i$ also solves the phase retrieval problem at all other frequencies if the corresponding measurement vectors $\mvec{b}_k =\matr{U}_{k,i} \mvec{b}_i$ are physically reasonable for every frequency $\omega_k$\,---\,at least if the inter-frequency mapping $\matr{U}_{k,i}$ is perfect.
 
Choosing an arbitrary reference frequency with index $i$, the multi-frequency phase retrieval problem is formulated as
\begin{align}
\text{find }\mvec{x}_i   \in \mathbb{C}^{N\times 1} \text{ such that }
\begin{pmatrix} 
\abs{\mvec{b}_1}\\
\vdots \\
\abs{\mvec{b}_i}\\
\vdots\\
\abs{\mvec{b}_{N_\mathrm{f}}}
\end{pmatrix} &= \abs{\, \begin{pmatrix}
	\matr{P}_1\,  \mvec{U}_{1,i} \, \matr{A}_i \\
	\vdots \\
	\matr{A}_i\\
	\vdots \\
	\matr{P}_{N_\mathrm{f}} \, \mvec{U}_{{N_\mathrm{f}},i} \, \matr{A}_i
	\end{pmatrix} \mvec{x}_i\, }\notag \\[0.5cm]
&= \abs{\, \tilde{\matr{A}}\, \mvec{x}_i\, }\, . \label{eq:unstable}
\end{align} 

If we consider a single measurement frequency only, we impose $M$ (possibly linearly dependent) restrictions on the source vector $\mvec{x}_i$. Using all $N_\mathrm{f}$ measurement frequencies at once, one can increase the number of restrictions on the source vector by a factor of $N_\mathrm{f}$ and it is clear that phase retrieval algorithms can be successful with larger probability.

The influence of the projection matrices $\matr{P}_k$ depends on the number of \ac{dof} $N_{\mathrm{dof}}$ of the radiated fields and the number of measurement samples.
The projection matrix is designed to filter out unphysical contributions in the vector ${\mvec{b}_k}$ which may be introduced by assigning false phases to the measured magnitudes. 
Only if the number of measurement samples is sufficiently large, the projection matrix will have a null-space which allows to separate unphysical solutions. 
With insufficient measurements, any vector $\mvec{b}\in \mathbb{C}^{M\times1}$ can be generated by the sources and appears to be physical ($\matr{P}_k$ is the identity matrix in this case).
The number of measurement samples should, therefore, exceed the number of \ac{dof}, $N_\mathrm{dof}$, considerably (e.g., by a factor of 2 or 4) at the highest measured frequency to ensure that unphysical solutions can be detected at all frequencies.

\subsection{Numerically Improved Formulation}
The problem of a division by zero in~\eqref{eq:unstable} can be treated by scaling the measurement samples at the frequency $\omega_k$ with the magnitudes of the signal at the reference frequency $\omega_i$.
The diagonal matrix 
\begin{align}
\matr{B}= \fun{\mathrm{diag}}{\abs{\mvec{b}_i}}
\end{align}
contains the magnitudes of the measurement vector $\mvec{b}_i$ at the reference frequency $\omega_i$. 
The linear relationship between the sources $\mvec{x}_k$ and $\matr{B}\mvec{b}_k$ (i.e.,the scaled version of the measurement samples  at frequency $\omega_k$) is given by the scaled matrix $\mvec{B} \mvec{A}_k$.
The relationship between the scaled measurement vector at frequency $\omega_k$ and the measurement vector at the reference frequency $\omega_i$ is given by the new diagonal matrix
\begin{align}
\tilde{\matr{U}}_{k,i} = \matr{B}\matr{U}_{k,i}
\end{align}
and its elements can be found by
\begin{align}
\left[\tilde{\matr{U}}_{k,i}\right]_{\ell \ell}= \abs{\left[\mvec{b}_k\right]_\ell} \ee^{\, \jj \left(\left[\mvec{\phi}_k\right]_\ell-\left[\mvec{\phi}_i\right]_\ell\right)}\, . \label{eq:Umat}
\end{align}

A modified projection matrix $\tilde{\matr{P}}_k = \matr{A}_k \left(\matr{B} \matr{A}_k\right)^\dagger$ must now account for the magnitude scaling of the measurement vector entries which is achieved by taking the pseudo inverse of the scaled forward matrix $\matr{B}\matr{A}_k$. Finally, the stabilized version of the multi-frequency phase retrieval problem reads 
\begin{align}
\text{find }\mvec{x}_i  \in \mathbb{C}^{N\times 1} \text{ such that }
\begin{pmatrix} 
\abs{\mvec{b}_1}\\
\vdots \\
\abs{\mvec{b}_i}\\
\vdots\\
\abs{\mvec{b}_{N_\mathrm{f}}}
\end{pmatrix} &= \abs{\, \begin{pmatrix}
	\tilde{\matr{P}}_1\,  \tilde{\matr{U}}_{1,i} \, \matr{A}_i \\
	\vdots \\
	\matr{A}_i\\
	\vdots \\
	\tilde{\matr{P}}_{N_{\mathrm{f}}} \tilde{\matr{U}}_{{{N_\mathrm{f}}},i} \, \matr{A}_i
	\end{pmatrix} \mvec{x}_i\, }\notag \\[0.5cm]
&= \abs{\, \tilde{\matr{A}}\, \mvec{x}_i\, }\, . 
\label{eq:stab1}
\end{align}
No direct division by small values occurs in this formulation. 
The division is de facto transferred to the modified pseudo inverse which may be able to retrieve the stability of the problem by regularization properties.

Both problem formulations in~\eqref{eq:unstable} and \eqref{eq:stab1} have the form of a standard phase retrieval problem and can be tackled with established algorithms.
In the following sections, we demonstrate that utilizing measurements at different frequencies provides a reliable method for generating large numbers of independent measurement samples for the phase retrieval problem such that a stable phase retrieval should be possible with any reconstruction method. 
Furthermore, the usage of different frequencies allows to generate independent measurement samples even in the \ac{ff}, where other methods fail to generate new independent samples, because no new information is obtained by using different probes or changing the distance to the \ac{aut}.

\subsection{Formulation Without Specific Reference Frequency}
Finally, if the measured signal at the reference frequency $f_i$ has a low magnitude for certain measurement samples, the measured reference phase may have an arbitrary error. 
Although we did not see this effect to affect our measured or simulated data,  an alternative formulation of the minimization problem in terms of measurement sample unknowns instead of \ac{aut} source unknowns shall be given in this section for completeness.

For every measurement location, we can choose one of the signal samples at a certain frequency as our unknown for this location. 
The remaining signal samples at this measurement location are then already determined because the ratio $\left[\mvec{b}_k\right]_\ell / \left[\mvec{b}_i\right]_\ell$ is known for all frequency pairs. 
If the signal magnitude is very low for all frequencies, the reconstructed phase is irrelevant, as the measured value is approximately zero for all frequencies. 
In all other cases, we can pick the most convenient frequency sample at each location to form our vector of unknowns and ensure a physically reasonable measurement vector by using the projection matrices. 
We have 
\begin{align}
\text{find }\hat{\mvec{b}}  \in \mathbb{C}^{N\times 1} \text{ such that } 
\begin{pmatrix} 
\abs{\mvec{b}_1}\\
\vdots \\
\abs{\mvec{b}_i}\\
\vdots\\
\abs{\mvec{b}_{N_\mathrm{f}}}
\end{pmatrix} &= \abs{\, \begin{pmatrix}
	\tilde{\matr{P}}_1\,  \hat{\matr{U}}_{1}  \\
	\vdots \\
	\tilde{\matr{P}}_i\,  \hat{\matr{U}}_{i}\\
	\vdots \\
	\tilde{\matr{P}}_{N_{\mathrm{f}}} \hat{\matr{U}}_{{{N_\mathrm{f}}}} 
	\end{pmatrix} \hat{\mvec{b}}\, }\notag \\[0.5cm]
&= \abs{\, \hat{\matr{A}}\, \hat{\mvec{b}}\, }\, 
\label{eq:stab2}
\end{align}
where $\hat{\mvec{b}}$ is the vector of unknown measurement samples at the chosen frequencies for each measurement location and $\hat{\matr{U}}_{i}$ is analogous to~\eqref{eq:Umat}, but with a variable reference frequency instead of a fixed reference frequency for every measurement location.

The magnitudes of all measurement samples in all the considered examples were well above any values which could be problematic for a numerical evaluation. Therefore, we did not see differences between the formulations with and without a fixed reference frequency. 
Since the newly included formulation introduces more unknowns in the reconstruction process, the originally proposed version remains the preferable choice and was used in the paper.
\section{Choosing the Frequency Step Size}\label{sec:step}
Our goal in this section is to find a rule of thumb for a reasonable frequency sample step size.
The received signal at a measurement position varies over frequency due to two different possible reasons for the change.
Either the current distribution on the \ac{aut} or the operator $\matr{A}_k$ changes.
A prediction of the frequency behavior of the \ac{aut} current distribution is difficult, hence we investigate the frequency behavior of the forward operator $\matr{A}$\footnote{When the current distribution changes, the proposed method will always generate independent data. Thus, a detailed investigation of changing currents is superfluous.}.

\begin{figure}
\centering
\includegraphics{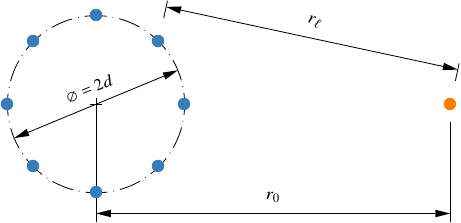}
	\caption{Setup to find a reasonable frequency step width.}\label{fig:freq}
\end{figure}

To this end, consider a simple example of $N$ $z$-oriented Hertzian dipoles placed on a ring with radius $d$ lying in the $xy$-plane as depicted in Fig.~\ref{fig:freq}.
These Hertzian dipoles represent a generic \ac{aut}. 
The field is sampled at a distance $r_0$ from the center of the dipole ring.

For the linear relationship between the sources and the $z$-component of the electric field at the observation location, we can write
\begin{equation}
\fun{E_z}{\omega}= \fun{\mvec{a}\herm}{\omega} \fun{\mvec{z}}{\omega}\, .
\end{equation}
For each frequency $\omega$, one obtains a row vector $\fun{\mvec{a}\herm}{\omega}$ describing the relationship between the sources and the observation point.

To provide new information, two row vectors $\fun{\mvec{a}\herm}{\omega_1}$ and $\fun{\mvec{a}\herm}{\omega_2}$ at different frequencies should not be too similar. 
As a similarity metric we can use the \ac{svd} of the matrix
\begin{equation}
\matr{A}= \begin{pmatrix}
\dfrac{1}{\omega_1}\fun{\mvec{a}\herm}{\omega_1} \\[1em]
\dfrac{1}{\omega_2}\fun{\mvec{a}\herm}{\omega_2}
\end{pmatrix}\, .
\end{equation}
The scaling with the inverse of the frequencies ensures, that the elementary dipole fields maintain the same field magnitude at different frequencies and that the two vectors $\fun{\mvec{a}\herm}{\omega_1}$ and $\fun{\mvec{a}\herm}{\omega_2}$ have their mean values on the same order of magnitude.
If the two row vectors are similar, the ratio $\abs{\sigma_2/\sigma_1}$ of the two singular values will be small.
If the two vectors are independent, then the ratio will approach unity.

\begin{figure}[t]
\centering
	\includegraphics{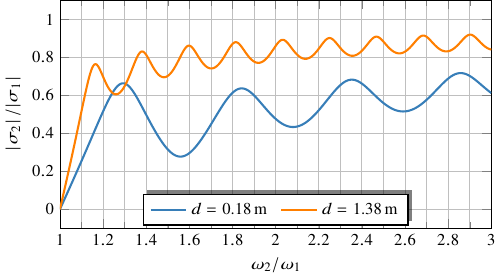}
	\caption{Singular value ratio for two different \ac{aut} sizes dependent on the frequencies of the forward operators with $\omega_1=2\uppi \, \SI{1}{GHz}$.} \label{fig:freq_res}
\end{figure}

Fig.~\ref{fig:freq_res} shows the singular value ratio for two different radii $d=\SI{0.18}{m}$ and $d=\SI{1.38}{m}$ of the \ac{aut} circle on which $N=1000$ dipoles were distributed evenly. The observation distance is $r_0=\SI{2.1}{m}$ and the frequency $\omega_1$ is fixed at $\omega_1=2\uppi \, \SI{1}{GHz}$.

Clearly, for larger antennas a smaller frequency step size is sufficient to ensure that the observation at the two frequencies will bring new information.
Overall, as soon as the \ac{aut} size exceeds the wavelength associated with the frequency step $\Delta \omega= \omega_2-\omega_1$ by some amount, we can expect that the forward operators for the two frequencies will show enough variation even if the current distribution on the \ac{aut} does not vary over the different frequencies.

Therefore, to maximize the information we can obtain from measurements in a certain bandwidth, the frequency step $\Delta \omega$ should not be larger than $\Delta \omega_\mathrm{max}=c_0/\left(2 d\right)$, where $d$ denotes the radius of the minimum sphere enclosing the \ac{aut} and $c_0$ is the speed of light.
If the frequency step size becomes much smaller than this value, however, many of the frequency samples will be redundant.   
 
\section{Results}\label{sec:results}
\subsection{Investigation on the Number of Independent Samples}  
For a first impression of the usefulness of the proposed multi-frequency formulation, the suitability of the operator $\tilde{\matr{A}}$ from \eqref{eq:stab1} for the task of phase retrieval is evaluated. 
Based on the technique described in~\cite{knapp_reconstruction_2019}, one can count, how many \enquote{independent} magnitude samples are possible with a given operator $\matr{A}$ by applying an \ac{svd} of the matrix
\begin{equation}
\matr{Q} = \abs{\matr{A\ \matr{A}\herm}}^{\circ 2}\, ,
\end{equation}
where $\matr{A}^{\circ 2}$ squares each element of $\matr{A}$.
The number of independent magnitude samples is then equal to the number of singular values with a magnitude larger than a given threshold~\cite{knapp_reconstruction_2019}.  
Larger values of this quantity correspond to an increased chance of successfully solving~\eqref{eq:phaseless_formulation}. 

For the purpose of comparison, we generated different linear operators $\matr{A}\in \mathbb{C}^{M\times N}$ for these unknowns.
The reference case consists of a matrix $\matr{A}_{\mathrm{Gaussian}}$, with i.i.d. Gaussian distributed elements.
Such a linear operator states the ideal case for a measurement matrix as each new measurement sample (i.e., each new row in $\matr{A}_{\mathrm{Gaussian}}$) provides independent information.
Once $N^2$ independent measurement samples have been reached, the phase retrieval problem becomes linear and every additional measurement sample will be (linearly) dependent on the previous samples.

Practical measurement matrices $\matr{A}$ are usually not Gaussian distributed. 
To investigate the behavior of measurement matrices in more realistic measurement scenarios, field measurements on either one sphere (radius $r_1 = 5\lambda$) or two spheres (radii $r_1 = 5 \lambda$, $r_2 = 10\lambda$) are considered.
The measurement locations are distributed on a Fibonacci spiral~\cite{Keinert.2015} on the measurement spheres, respectively.
This ensures that the $M$ measurement locations are relatively evenly distributed on the surfaces for varying values of $M$.
Two measurement samples are taken at every measurement location, corresponding to the two orthogonal polarizations.

A synthetic spherical measurement setup is considered. The \ac{aut} is modeled by $N=300$ randomly excited Hertzian dipoles distributed according to a Fibonacci mapping~\cite{Keinert.2015} tangentially on the surface of a sphere with diameter of $2\lambda$.
The random excitation of the dipoles are the same for all frequencies, such that variations in the matrix $\tilde{\matr{A}}$ solely come from variations of the forward operator at different frequencies.  

On the single measurement sphere, we also assume a probe array similar to~\cite{paulus_phaseless_2017}, which can perform (complex valued) linear combinations of two separated field values before the magnitude of the signal is taken. The separation between the two field samples was set to $2\lambda$. 
The special probe antennas yield  $8$ measurements, consisting of four linear combinations for each of the combinations of two polarizations between the two measurement locations of the array, where the displacement of the array elements can be either along the $\vartheta$- or the $\varphi$-direction. 
From the magnitudes of these linear combinations, local phase information in the form of phase differences can be obtained~\cite{paulus_phaseless_2017,knapp_reconstruction_2019,costanzo_novel_2005, costanzo_wideband_2008}.

\begin{figure}[t]
\centering
	\includegraphics[width=0.49 \textwidth]{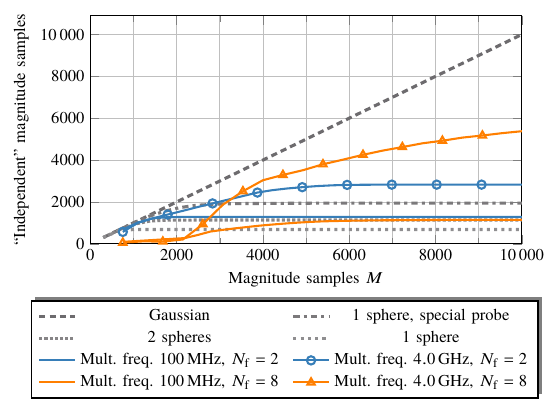}
	\caption{Number of “independent” magnitude samples for different linear operators corresponding to various \ac{nf} measurement setups.} \label{fig:fig1}
\end{figure}

Finally, the measurement matrix $\tilde{\matr{A}}$ is generated according to~\eqref{eq:stab1} from the measurement matrices $\matr{A}_k$ at multiple frequencies for measurement samples on the single measurement sphere with radius $r_1$.
For the sake of a fair comparison of the measurement setups, the number of rows, $M$, in the resulting forward operators is the same for each approach.
This means that fewer measurement locations are considered for the probe array and multi-frequency setups such that the number $M$ of actual measurement samples is the same for each setup.
The results can be seen in Fig.~\ref{fig:fig1}, where the number of frequencies utilized in the multi-frequency formulation was set to $N_\mathrm{f} = 2$ and $N_\mathrm{f} = 8$, respectively. 
For each case, two bandwidths were considered: a small bandwidth of $\SI{100}{\MHz}$ between $\SI{3.0}{\GHz}$ and $\SI{3.1}{\GHz}$, and a large bandwidth of $\SI{4}{\GHz}$ between $\SI{3.0}{\GHz}$ and $\SI{7.0}{\GHz}$.

For different numbers of samples $M$, we count the number of singular values $\sigma_i$ which have a ratio $\sigma_i/\sigma_1\geq10^{-5}$ to the largest singular value. 
As a matter of fact, the number of independent samples is limited to $N^2=9000$. 
While the Gaussian matrix shows an ideal behavior, where the number of independent samples rises linearly with every additional sample until the upper limit is reached at $M=9000$, the curves for the more realistic cases saturate on certain levels. 
The probe array and the multi-frequency formulation with the larger bandwidth may yield the best results for large oversampling in the considered scenario.
This suggests that more valuable information can be collected by using specialized probe arrays or the broadband information on a single sphere, as compared to measurements on two spheres with the same probe at the same frequency.

The multi-frequency formulation with the small bandwidth stagnates at a level only slightly larger than the measurements at one frequency. 
The measurement matrices at neighboring frequencies do not differ much from each other and the rows of the combined operator $\tilde{\matr{A}}$ are correlated.

If the number of measurement locations is too small, the co-kernel of the matrix $\matr{A}$ vanishes. 
The projection matrices $\matr{P}_k=\matr{A}_k^{ }\matr{A}_k^\dagger$ collapse to the unit matrix and have no influence.
Once the number of measurement locations exceeds $N_\mathrm{dof}$ for the antenna model at the corresponding frequency, the projection matrices become effective.

The measurement samples at different frequencies only have an influence, if enough spatial sampling locations are considered.
When the spatial sampling exceeds the critical value, all of a sudden, the filtering properties of the projection matrices have an effect and the measured signal samples at the different frequencies convey independent information.
Therefore, the number of independent samples for the $N_\mathrm{f}=8$ cases grow very slowly at first (only about every eighth sample is useful) before we have a drastic increase of independent samples (slope greater than one) once the critical sampling rates for the antenna model at the respective frequencies are reached.

\begin{figure}[t]
\centering
	\includegraphics[width=0.49 \textwidth]{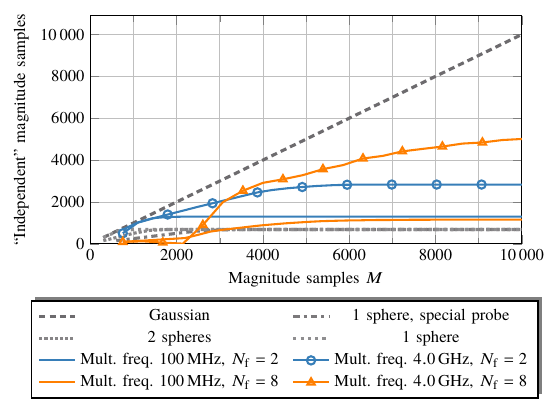}
	\caption{Number of “independent” magnitude samples for different linear operators corresponding to various \ac{ff} measurement setups.} \label{fig:fig2}
\end{figure}

In Fig.~\ref{fig:fig2}, the same analysis is performed in the \ac{ff} ($r_1=\num{5000} \lambda$, $r_2=\num{10000} \lambda$).
It can be seen that measurements at two distances or with special probes are not able to generate more independent measurements than a simple probe at a single \ac{ff} distance.
Only the multi-frequency measurements \emph{are} indeed able to increase the number of independent samples to a level which is comparable to the \ac{nf} measurements.
\subsection{Field Transformation of a Simulated Biconical Antenna}
\begin{figure}[t]
\centering
		\includegraphics[width = 5cm]{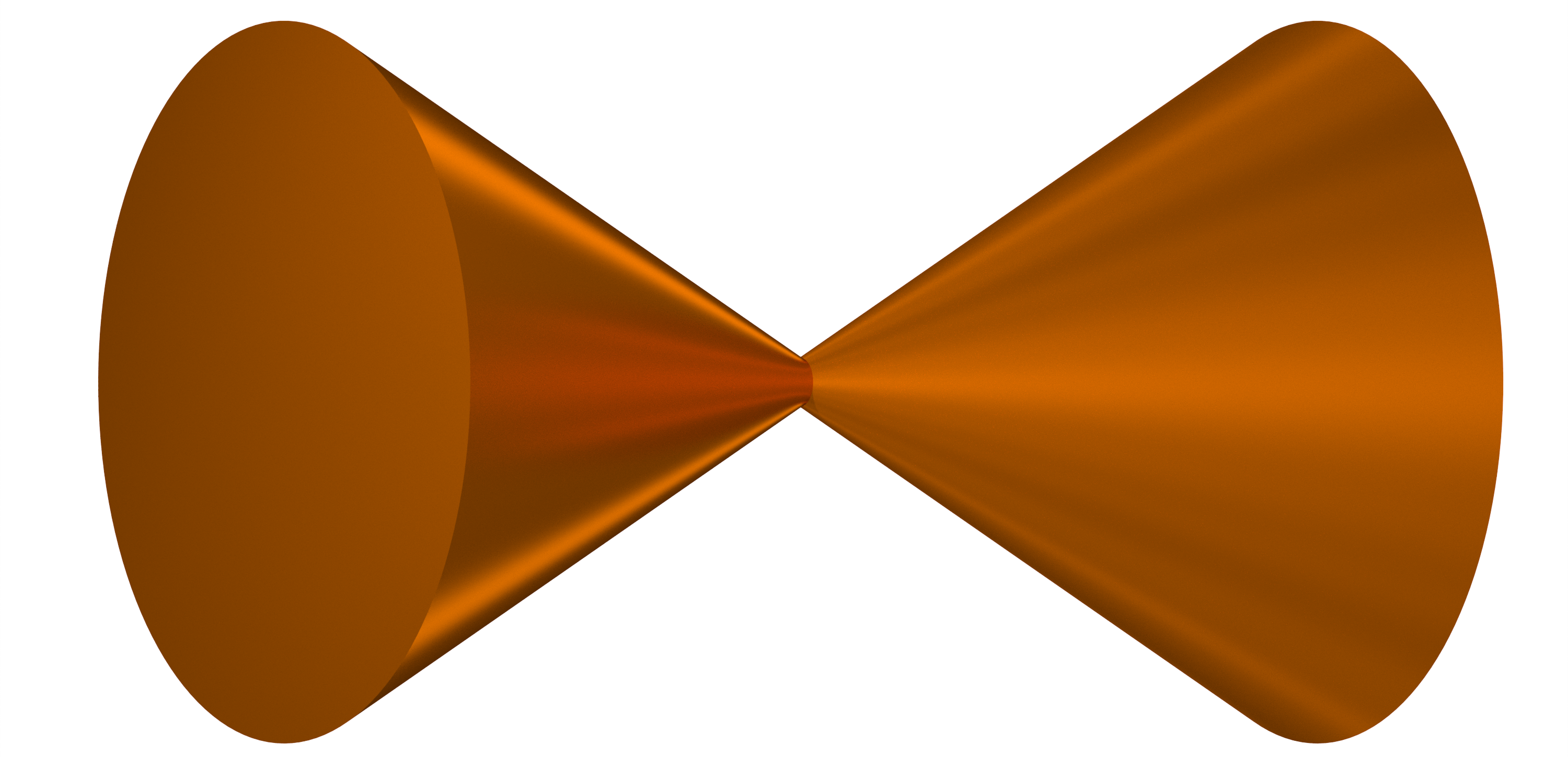}	
	\caption{Simulated biconical dipole antenna.} \label{fig:biconical}
\end{figure}

In this example, we consider a biconical antenna as depicted in Fig.~\ref{fig:biconical}. 
The biconical antenna with a minimum sphere diameter of around \SI{0.33}{m} was simulated with Feko~\cite{altair_feko_} in a frequency range from \SI{2}{GHz} to \SI{4}{GHz}.
The \ac{nf} was sampled on three spherical surfaces, with radii $r_1 = \SI{0.9}{\meter}$, $r_2 = \SI{1.2}{\meter}$ and $r_3 = \SI{1.5}{\meter}$.
On every surface the two tangential field components were obtained at $\num{8000}$ locations.

For the \ac{ff} reconstruction, the equivalent source model consists of \num{3001} Hertzian dipoles distributed on a hull surface enclosing the \ac{aut}.
The excitation coefficients of the Hertzian dipoles were obtained by solving~\eqref{eq:stab1} in a Wirtinger-Flow-like minimization procedure~\cite{paulus_phaseless_2017,candes_phase_2015}.
After the source reconstruction, the corresponding \ac{ff} is obtained from the reconstructed sources.

\begin{figure}[p]
	\centering
	\includegraphics{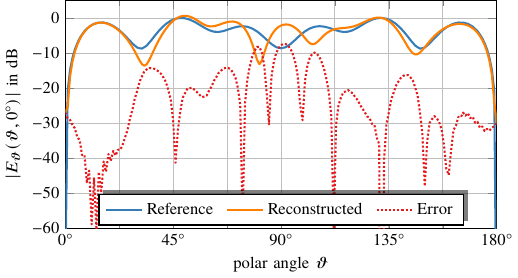} 
	\caption{Reconstructed $\varphi=0^\circ$ \ac{ff}-cut of the $\vartheta$-component at \SI{2}{GHz} using three measurement distances and a single frequency.}\label{fig:bic2GHz_2}
\end{figure}

\begin{figure}[p]
	\centering
	\includegraphics{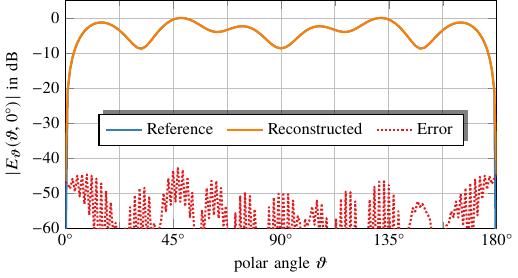} 
	\caption{Reconstructed $\varphi=0^\circ$ \ac{ff}-cut of the $\vartheta$-component at \SI{2}{GHz} using three measurement distances and eleven distinct frequencies.}\label{fig:bic2GHz_4}
\end{figure}

\begin{figure}[tp]
	\centering
	\includegraphics{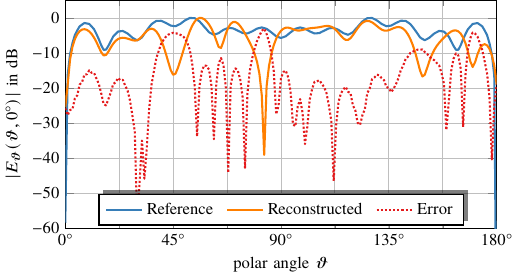} 
	\caption{Reconstructed $\varphi=0^\circ$ \ac{ff}-cut of the $\vartheta$-component at \SI{4}{GHz} using three measurement distances and a single frequency.}\label{fig:bic4GHz_2}
\end{figure}

\begin{figure}[tp]
	\centering
	\includegraphics{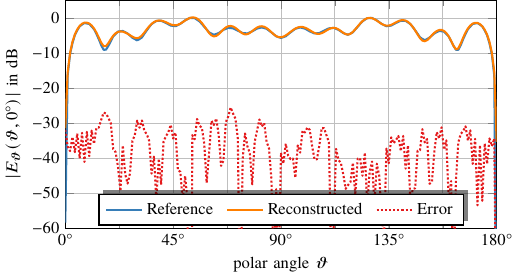} 
	\caption{Reconstructed $\varphi=0^\circ$ \ac{ff}-cut of the $\vartheta$-component at \SI{4}{GHz} using three measurement distances and eleven frequencies.}\label{fig:bic4GHz_4}
\end{figure}

Figs.~\ref{fig:bic2GHz_2} to \ref{fig:bic4GHz_4} show the reconstructed \acp{ff} at \SI{2}{GHz} and \SI{4}{GHz} using either a single or eleven different frequencies in $\tilde{\matr{A}}$, respectively.
All \acp{ff} are normalized to their respective maximum.
The error curve is given by
\begin{equation}
\fun{\epsilon_\mathrm{dB}}{\vartheta,\varphi}= 20 \fun{\log_{10}}{\abs{\dfrac{\abs{\fun{E_\vartheta}{\vartheta,\varphi}}}{E_{\vartheta,\mathrm{max}}}-\dfrac{\abs{\fun{E_{\vartheta,\mathrm{ref}}}{\vartheta,\varphi}}}{E_{\vartheta,\mathrm{max},\mathrm{ref}}}}}\,.
\end{equation}

Adding more frequency samples helps to avoid getting stuck in local minima during the solution of~\eqref{eq:stab1}. 
Using only a single frequency at a time is not enough to reconstruct a satisfying solution.
Adding additional restrictions in form of measurements at additional frequencies leads to an accurate \ac{ff} reconstruction.
Naturally, since the relative phases between signal samples at the different frequencies at every measurement location are known, one obtains the correct solution at all frequencies, as soon as the correct solution has been found for one frequency.

\newpage
\subsection{Field Transformation of Horn Antenna Measurements}\label{sec:hornmeasurement}
\subsubsection{The Measurement Setup}
\begin{figure}[t]\centering
		\includegraphics[width=0.4 \textwidth]{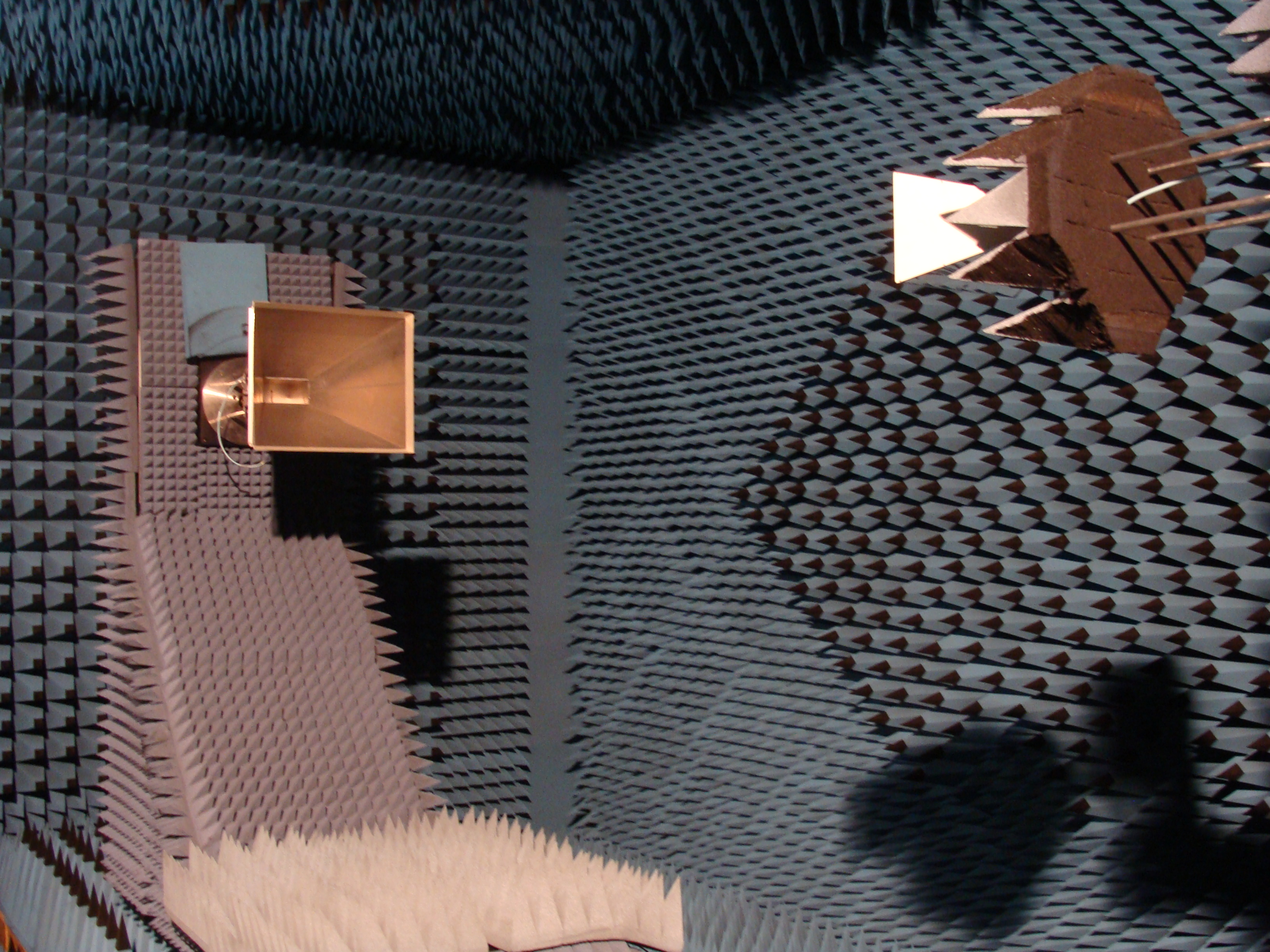} 
	\caption{Measurement setup with DRH18 probe and the horn \ac{aut}.}\label{fig:setup}
\end{figure}
\ac{nf} measurements were obtained in the anechoic antenna measurement chamber at the Technical University of Munich, demonstrating the feasibility of the approach for realistic noisy data.
The radiated field of a horn antenna with minimum sphere radius of about \SI{0.42}{m} was sampled with a DRH18 probe~\cite{rfspin} on four spherical surfaces with radii $r_1=\SI{2.512}{m}$, $r_2=\SI{2.642}{m}$, $r_3=\SI{2.892}{m}$, and $r_4=\SI{3.092}{m}$ 
at a large number of frequencies in $25$\,MHz steps.
On each of the measurement surfaces, \num{130682} measurement samples were obtained coherently with magnitude and phase on a regular \SI{1}{\degree}-grid\,---\,this represents a strong oversampling of the field, $N_\mathrm{dof}$ is on the order of several thousands for an antenna of this size at around \SI{3}{GHz}. 
Fig.~\ref{fig:setup} shows the \ac{aut} horn antenna and the DRH18 probe antenna mounted in the anechoic chamber. 

From the large number of measurement samples, \num{20000} samples were picked to reduce the computational complexity.
A reference \ac{ff} from the measured \ac{nf} data was calculated with a conventional \ac{nf} to \ac{ff} transformation from the coherent data.
The phase information was then removed from the \ac{nf} data and knowledge of phase differences between the frequency samples was introduced in the form of~\eqref{eq:stab1}.
\num{5000} Hertzian dipoles placed on a conformal hull around the \ac{aut} served as the \ac{aut} model for the transformations.

The single frequency phase-retrieval problem is solved for each individual frequency with a non-convex solver~\cite{paulus_phaseless_2017}, where an initial guess according to a spectral method~\cite{candes_phase_2015} for the matrix $\matr{A}_k$ is employed.
The procedure for the multi-frequency transformations is as follows. 
First, we solve the single frequency phaseless problem for the reference frequency as mentioned above.
This solution $\mvec x _i$ is taken as an initial guess for the multi-frequency phase retrieval problem.
The solution of the multi-frequency problem with the non-convex solver gives an estimate of the phases of the measurement vectors at each frequency. 
Then, we obtain complex measurement vectors by combining the estimated phases with the measured magnitude vectors.
We retrieve the equivalent currents at each frequency by solving a standard complex problem. 

\subsubsection{The Problem of Local Minima}
\begin{figure}[t]
	\centering
	\includegraphics{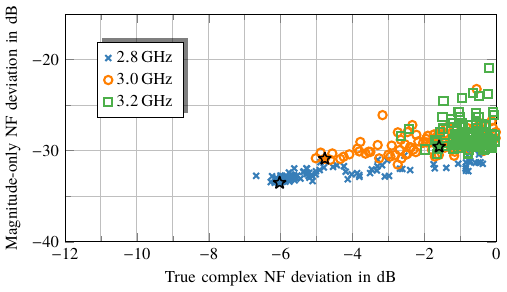} 
	\caption{Comparison of magnitude-only and complex \ac{nf} deviations for single-frequency measurements on a Fibonacci spiral and four distances. The star marks the initial guess by the spectral method, the other initial guesses are random.}\label{fig:scatter_single}
\end{figure}

\begin{figure}[t]
	\centering
	\includegraphics{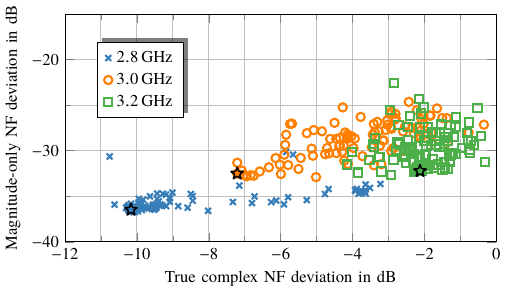} 
	\caption{Comparison of magnitude-only and complex \ac{nf} deviations for single-frequency measurements on a regular grid  and four distances. The star marks the initial guess by the spectral method, the other initial guesses are random.}\label{fig:scatter_single2}
\end{figure}

The initial guess has a large impact on the solution of the phase-retrieval problem.
A severe problem of non-convex phase retrieval techniques (i.e., the minimization of non-linear non-convex cost functions) are local minima. 
With ideal data, we can try to judge whether we are stuck in a local minimum\,---\,only the hopefully unique global minimum exhibits a cost function value of zero.
In measurement scenarios, the problem is worse: The measurement error gives a limit to the reconstruction deviation, and local minima with cost function values below the error floor become indistinguishable from the true solution. This is observed in Fig.~\ref{fig:scatter_single} for Fibonacci spiral sampling and in Fig.~\ref{fig:scatter_single2} for randomly picked samples from a regular grid with the pole in the main-beam direction.
For $100$ random right-hand sides, we investigate the solver behavior at $2.8$, $3.0$, and $3.2$\,GHz and observe that there is no way to know whether a lower magnitude-only deviation $\big\lVert\lvert\mvec b\rvert-\lvert\tilde{\matr{A}}\, \mvec{x}\rvert\big\rVert_2 $ correlates with a better true complex \ac{nf} deviation $\lVert\mvec b-\tilde{\matr{A}}\, \mvec{x}\rVert_2 $.
With the good initial guess provided by the mentioned spectral method (marked with a star), the magnitude-only solver is also stuck in a local minimum.
Many (possibly good or bad) solutions with similar magnitude-only \ac{nf} deviation exist. 
It is not possible to decide whether a good magnitude deviation implies a good solution since many of the local minima are indistinguishable for the phaseless solver.
Apparently, the four measurement surfaces did not provide enough diversity in the measured data to avoid local minima.

\begin{figure}[t]
	\includegraphics{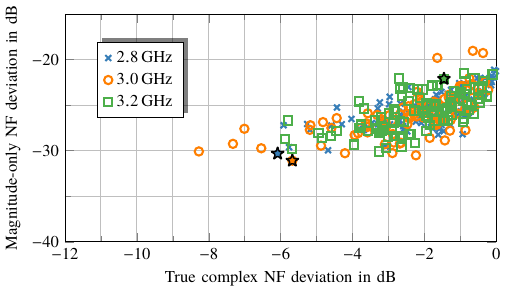} 
	\caption{Comparison of magnitude-only and complex \ac{nf} deviations for three-frequencies measurements on a Fibonacci spiral and four distances, three choices for the reference frequency. The star marks the initial guess from the single-frequency solution, the other initial guesses are random.}\label{fig:scattermulti}
\end{figure}

\begin{figure}[t]
	\includegraphics{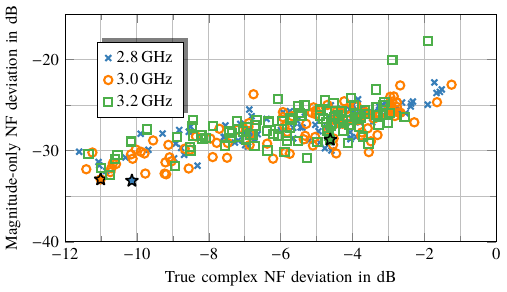} 
	\caption{Comparison of magnitude-only and complex \ac{nf} deviations for three-frequencies measurements on a regular sampling and four distances, three choices for the reference frequency. The star marks the initial guess from the single-frequency solution, the other initial guesses are random.}\label{fig:scattermulti2}
\end{figure}

An advantage of the multi-frequency method is found in the fact that the quality of the solution can be judged to some extent by looking at the reconstruction deviation of the multi-frequency solution. With the discussed measurement data, we can also compare the magnitude reconstruction deviation to the true complex \ac{nf} deviation. The result is shown in Figs.~\ref{fig:scattermulti} and~\ref{fig:scattermulti2} (for Fibonacci and regular sampling) for the multi-frequency case with three frequencies, where each of them is chosen as reference in~\eqref{eq:stab1}.
We observe that local minima with rather low magnitude deviation are still present, but most of the local minima are shifted to much larger \ac{nf} reconstruction deviation values.

Analyzing the accuracy of the different variants, we see that regular sampling performs here better than spiral sampling.

\subsubsection{Transformed FF Patterns}
\begin{figure}[tp]
\centering
	\includegraphics{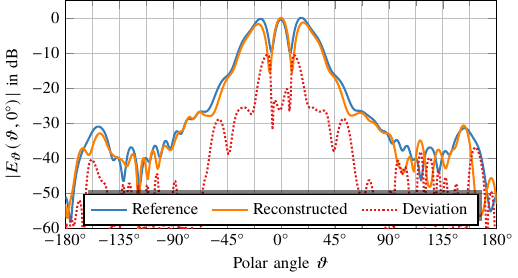} 
	\caption{Reconstructed $\varphi=0^\circ$ \ac{ff}-cut of the $\vartheta$-component at \SI{3.2}{GHz} using four measurement distances and a single frequency.}\label{fig:meas3p2GHztheta_single}
\end{figure}

\begin{figure}[tp]
\centering
	\includegraphics{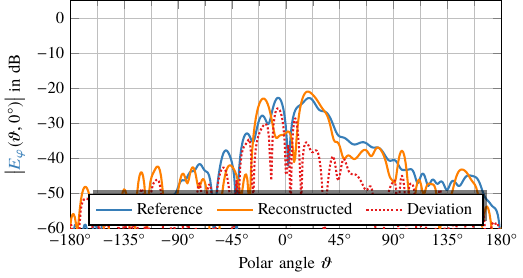} 
	\caption{Reconstructed $\varphi=0^\circ$ \ac{ff}-cut of the $\varphi$-component at \SI{3.2}{GHz} using four measurement distances and a single frequency.}\label{fig:meas3p2GHzphi_single}
\end{figure}

The \acp{ff} in Figs.~\ref{fig:meas3p2GHztheta_single} and \ref{fig:meas3p2GHzphi_single} were obtained from the magnitudes at \SI{3.2}{GHz} only.
\begin{figure}[tp]
\centering
	\includegraphics{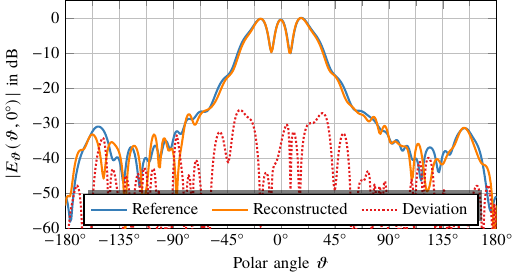} 
	\caption{Reconstructed $\varphi=0^\circ$ \ac{ff}-cut of the $\vartheta$-component at \SI{3.2}{GHz} using four measurement distances and three frequencies.}\label{fig:meas3p2GHztheta_mult}
\end{figure}
\begin{figure}[tp]
\centering
	\includegraphics{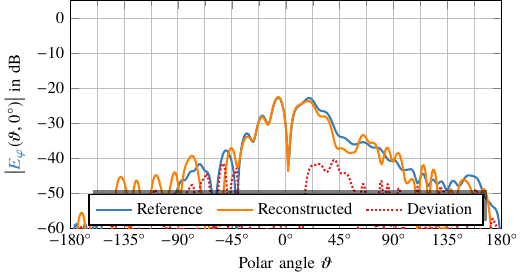} 
	\caption{Reconstructed $\varphi=0^\circ$ \ac{ff}-cut of the $\varphi$-component at \SI{3.2}{GHz} using four measurement distances and three frequencies.}\label{fig:meas3p2GHzphi_mult}
\end{figure}
Using only a single frequency, the phase reconstruction fails despite having measurements on four distances to the \ac{aut}.
The magnitude-only NF deviations and the complex NF deviations are shown in Fig.~\ref{fig:scatter_single2}.
The maximum \ac{ff} errors for the frequencies $2.8$, $3.0$, and $3.2$\,GHz are $-23.0$, $-19.6$, and $-10.3$\,dB, respectively.

Figs.~\ref{fig:meas3p2GHztheta_mult} to \ref{fig:meas3p2GHzphi_mult} show the reconstructed \acp{ff} for the $\vartheta$- and $\varphi$- components of the electric field at \SI{3.2}{GHz}, where the knowledge of the phase differences related to the three frequencies is used in the manner of~\eqref{eq:stab1} with \SI{2.8}{GHz} as reference frequency.   
The additional information obtained by using the three frequencies helped in this case to find a much better solution in the minimization process.
The NF reconstruction deviations for the combined problem are given in Fig.~\ref{fig:scattermulti2}, the subsequent solution of the  individual problems yield NF magnitude deviations of $-35.5$, $-31.1$, and $-33.6$\,dB, which correspond to NF errors of $-10.4$, $-9.2$, and $-10.2$\,dB.
The reconstructed \ac{ff} is in agreement with the reference \ac{ff} up to a maximum deviation of $-24.0$, $-24.9$, and $-26.1$\,dB for the three investigated frequencies\,---\,which is an improvement at each frequency.

\begin{figure}[tp]
\centering
	\includegraphics{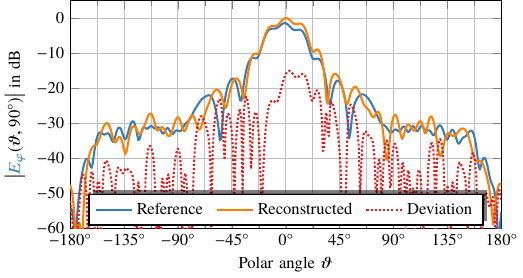} 
	\caption{Reconstructed $\varphi=90^\circ$ \ac{ff}-cut of the $\varphi$-component at \SI{3.6}{GHz} using four measurement distances and a single frequency.}\label{fig:meas3p6GHzphi_single}
\end{figure}

\begin{figure}[tp]
\centering
	\includegraphics{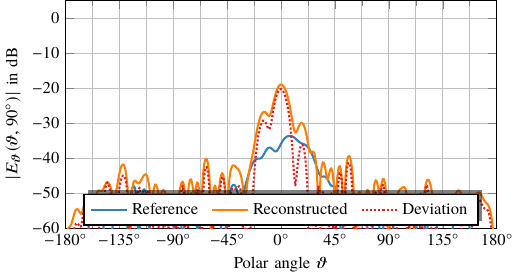} 
	\caption{Reconstructed $\varphi=90^\circ$ \ac{ff}-cut of the $\vartheta$-component at \SI{3.6}{GHz} using four measurement distances and a single frequency.}\label{fig:meas3p6GHztheta_single}
\end{figure}

\begin{figure}[tp]
\centering
\includegraphics{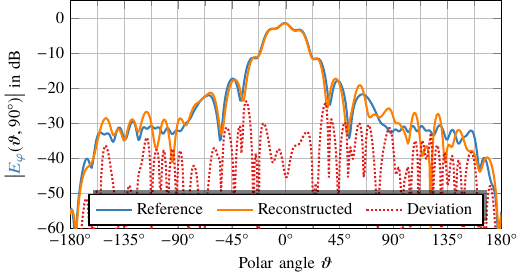}
\caption{Reconstructed $\varphi=90^\circ$ \ac{ff}-cut of the $\varphi$-component at \SI{3.6}{GHz} using four measurement distances and nine frequencies.}\label{fig:meas3p6GHzphi_mult}
\end{figure}

\begin{figure}[tp]
\centering
	\includegraphics{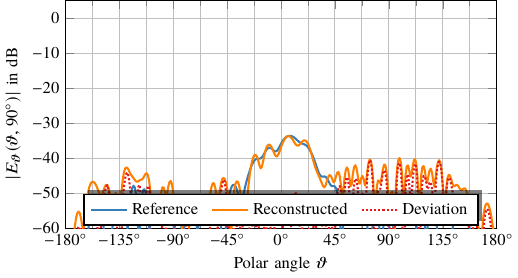}
	\caption{Reconstructed $\varphi=90^\circ$ \ac{ff}-cut of the $\vartheta$-component at \SI{3.6}{GHz} using four measurement distances and nine frequencies.}\label{fig:meas3p6GHztheta_mult}
\end{figure}

The method also works for larger bandwidths. We performed the same procedure for the range $2.8$ to $3.6$\,GHz with $100$\,MHz steps.
NF and FF error improvements are similar as in the discussed case with three frequencies. In particular, the FF patterns of the single-frequency and multi-frequency cases are shown in Figs.~\ref{fig:meas3p6GHzphi_single} to~\ref{fig:meas3p6GHztheta_mult}.
Note that this time the shown frequency is \SI{3.6}{GHz} and the shown cut is $\varphi=90^\circ$. 
	The pattern maximum does not appear in this cut at this frequency.


\subsection{Planar Field Transformation of Synthetic Horn Antenna Measurements}\label{sec:hornplanar}

In order to verify that the procedure is not restricted to spherical surfaces, a planar measurement setup has been synthetically mimicked from the source distribution found for the horn antenna in the previous example. The $x$- and $y$-components of the field were computed on two planar surfaces of size $\SI{3}{m}\times\SI{3}{m}$ at distances \SI{1.5}{m} and \SI{1.7}{m}, yielding a valid angle for the \ac{nffft} of about $\pm 60^\circ$. The spatial sampling step was $\Delta x= \Delta y=\SI{0.03}{m}$.  The calculated field magnitudes and the phase differences between the frequency components at 9 frequencies from \SI{2.8}{GHz} to \SI{3.6}{GHz} were taken as input for the multi-frequency algorithm.  

\begin{figure}[tp]
	\centering
	\includegraphics{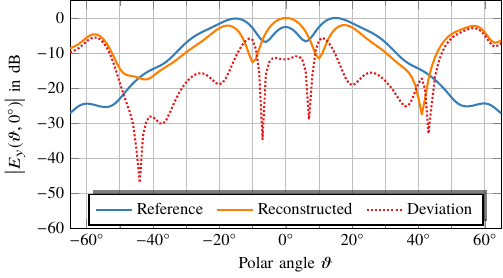}
	\caption{Reconstructed $\varphi=0^\circ$ \ac{ff}-cut of the $y$-component at \SI{3.5}{GHz} using two measurement distances and a single frequency.}\label{fig:2planes_single}
\end{figure}
\begin{figure}[tp]
	\centering
	\includegraphics{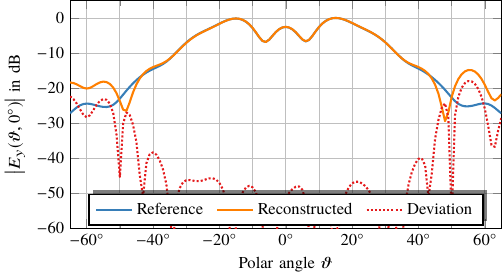}
	\caption{Reconstructed $\varphi=0^\circ$ \ac{ff}-cut of the $y$-component at \SI{3.5}{GHz} using two measurement distances and nine frequencies.}\label{fig:2planes_mult}
\end{figure}
The reconstructed co-polar components of the \acp{ff} for the single frequency phase reconstruction algorithm in Fig.~\ref{fig:2planes_single} and the multi-frequency algorithm in Fig.~\ref{fig:2planes_mult} show that the reconstruction algorithm can benefit from multi-frequency data, regardless of the particular sampling geometry.
  \begin{table}[tp]
	\centering
	\footnotesize
	\caption{Mean \ac{nf}-reconstruction errors}
	\label{tab:errors}
	\begin{tabular}{ l r r r r}
		\toprule
		 & 
		\multicolumn{2}{c}{Single-frequency} &
		\multicolumn{2}{c}{Multi-frequency} \\
		\cmidrule(lr){2-3}\cmidrule(l){4-5}
		Frequency &
		\multicolumn{1}{c}{$\epsilon_{\mathrm{mag}}$} & 
		\multicolumn{1}{c}{$\epsilon_{\mathrm{compl}}$} &
		\multicolumn{1}{c}{$\epsilon_{\mathrm{mag}}$} & 
		\multicolumn{1}{c}{$\epsilon_{\mathrm{compl}}$} \\
		\cmidrule(r){1-1}\cmidrule(lr){2-2}\cmidrule(lr){3-3}\cmidrule(lr){4-4}
		\cmidrule(l){5-5}
		\SI{2.8}{GHz} & \SI{-52.3}{dB} & \SI{-11.3}{dB}& \SI{-50.8}{dB} & \SI{-10.6}{dB}\\
		\SI{2.9}{GHz} & \SI{-47.7}{dB} & \SI{-17.9}{dB}& \SI{-46.6}{dB} & \SI{-12.9}{dB}\\
		\SI{3.0}{GHz} & \SI{-36.2}{dB} & \SI{-7.3}{dB} & \SI{-47.4}{dB} & \SI{-11.8}{dB}\\
		\SI{3.1}{GHz} & \SI{-33.0}{dB} & \SI{-2.7}{dB} & \SI{-50.2}{dB} & \SI{-12.5}{dB}\\
		\SI{3.2}{GHz} & \SI{-35.1}{dB} & \SI{-1.9}{dB} & \SI{-49.2}{dB} & \SI{-12.4}{dB}\\
		\SI{3.3}{GHz} & \SI{-35.2}{dB} & \SI{-2.3}{dB} & \SI{-50.7}{dB} & \SI{-12.8}{dB}\\
		\SI{3.4}{GHz} & \SI{-33.8}{dB} & \SI{-1.7}{dB} & \SI{-49.5}{dB} & \SI{-13.0}{dB}\\
		\SI{3.5}{GHz} & \SI{-33.0}{dB} & \SI{-0.7}{dB} & \SI{-50.8}{dB} & \SI{-13.3}{dB}\\
		\SI{3.6}{GHz} & \SI{-47.4}{dB} & \SI{-13.2}{dB}& \SI{-51.6}{dB} & \SI{-11.6}{dB}\\
		\bottomrule
	\end{tabular}
\end{table}
Table~\ref{tab:errors} shows the mean magnitude reconstruction error $\epsilon_{\mathrm{mag}}= \norm{\abs{\vec{b}}-\abs{\vec{b}_\mathrm{ref}}}/\norm{\abs{\vec{b}_\mathrm{ref}}}$ and the mean complex reconstruction error $\epsilon_{\mathrm{compl}}= \norm{{\vec{b}}-{\vec{b}_\mathrm{ref}}}/\norm{{\vec{b}_\mathrm{ref}}}$ for the single-frequency phase and the multi-frequency reconstruction algorithm. A comparison reveals that the magnitude must be very accurately matched by the reconstruction algorithms in order to obtain acceptable phase reconstructions and that using multi-frequency data evidently increases the probability of a correct reconstruction.
\section{Conclusion}

In general, phase retrieval for antenna measurements is an extremely hard to solve problem, since it is non-linear and non-convex. Given that only one correct and unique global minimum exists (which cannot be ensured in
practice due to measurement errors), there are no reliable algorithms available to find this global minimum
with reasonable effort. Any piece of additional information helps the algorithms to come (statistically) closer
to this minimum\,---\,where it has to be clear that this is by no means a guarantee for a specific set of measurements.

Using the relative phase differences of signal samples at different frequencies at the same measurement location can help to improve the reliability of the non-convex phase retrieval algorithms in antenna measurements.
The relative phase information can be measured and the corresponding measurement setup is less challenging than a fully coherent measurement setup at very high frequencies. 
Having in mind that every available piece of information is useful, we know that measurements on multiple surfaces are (statistically) superior to measurements on a subset of these surfaces. The same
holds for adding multi-frequency data: It helps to make existing methods (e.g., with multiple measurement surfaces)
more reliable.

In our investigations, the multi-frequency formulation was found to be able to achieve better results than a single frequency formulation for otherwise unchanged measurement configurations, whereas we never faced the case in which the single-frequency solution was better.

\ifCLASSOPTIONcaptionsoff
  \newpage
\fi

\bibliographystyle{IEEEtran}
\bibliography{multifrequency}
	\begin{IEEEbiography}[{\includegraphics[width=1in,height=1.25in,clip,keepaspectratio]{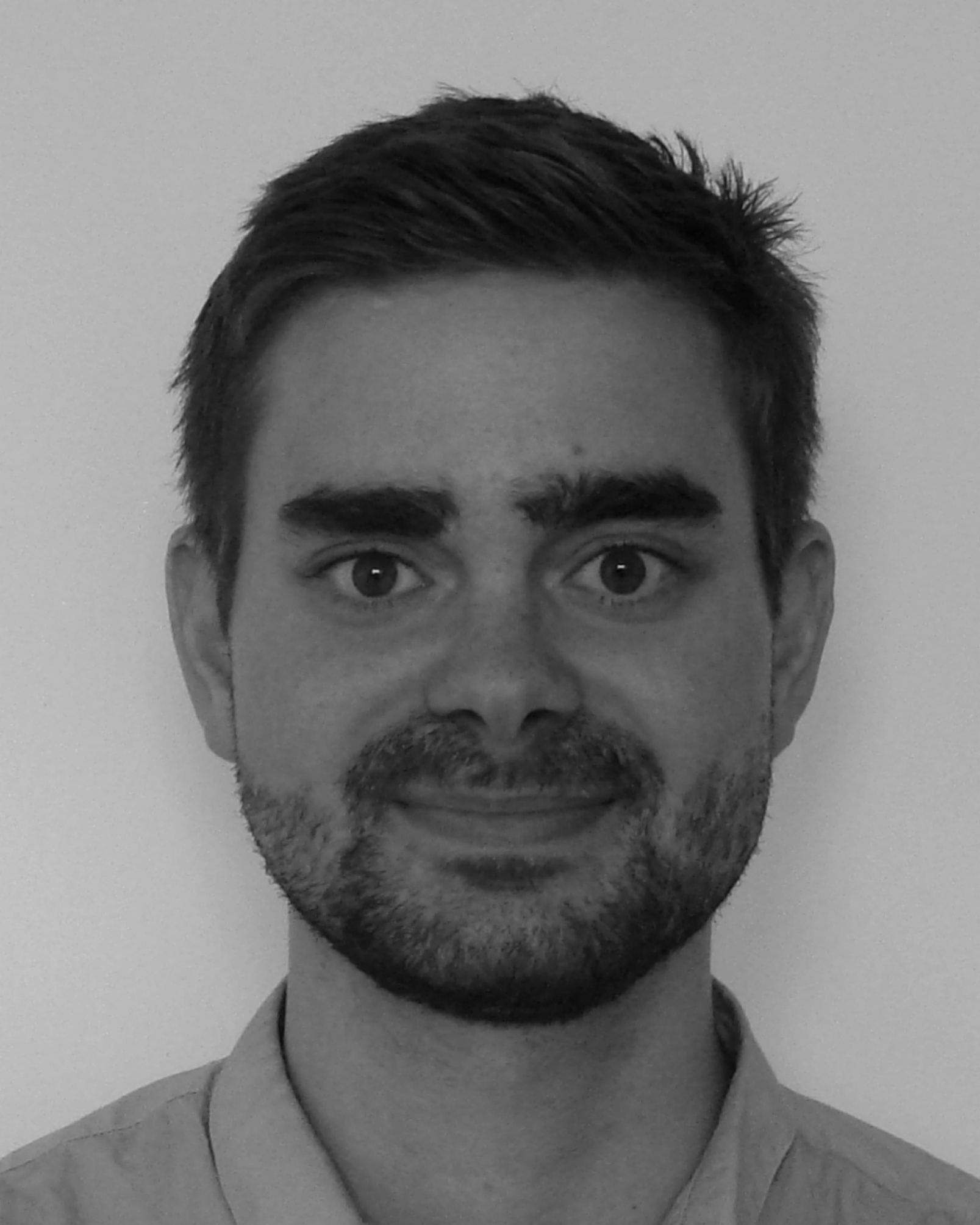}}]{Josef Knapp}
	
	(S'18) received the M.Sc. degree in electrical engineering and information technology from the Technical University of Munich, Munich, Germany, in 2016. Since 2016, he has been a Research Assistant at the Chair of High-Frequency Engineering, Department of Electrical and Computer Engineering, Technical University of Munich. His research interests include inverse electromagnetic problems, computational electromagnetics, antenna measurement  techniques in unusual environments, and field transformation techniques.
\end{IEEEbiography}
\begin{IEEEbiography}[{\includegraphics[width=1in,height=1.25in,clip,keepaspectratio]{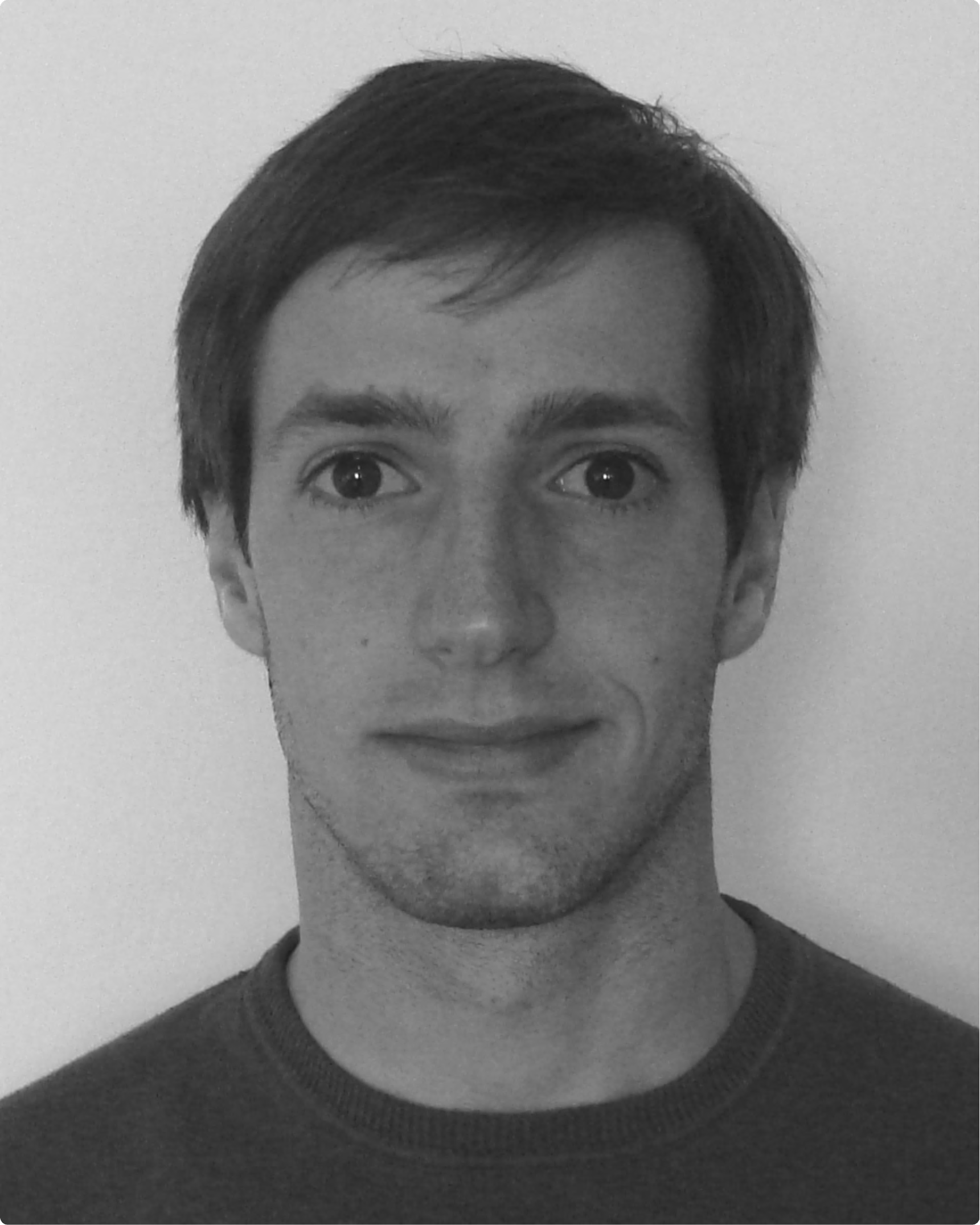}}]{Alexander Paulus}
	
	(S'18) received the M.Sc. degree in electrical engineering and information technology from the Technical University of Munich, Munich, Germany, in 2015. Since 2015, he has been a Research Assistant at the Chair of High-Frequency Engineering, Department of Electrical and Computer Engineering, Technical University of Munich. His research interests include inverse electromagnetic problems, computational electromagnetics and antenna measurement techniques.
\end{IEEEbiography}
\begin{IEEEbiography}[{\includegraphics[width=1in,height=1.25in,clip,keepaspectratio]{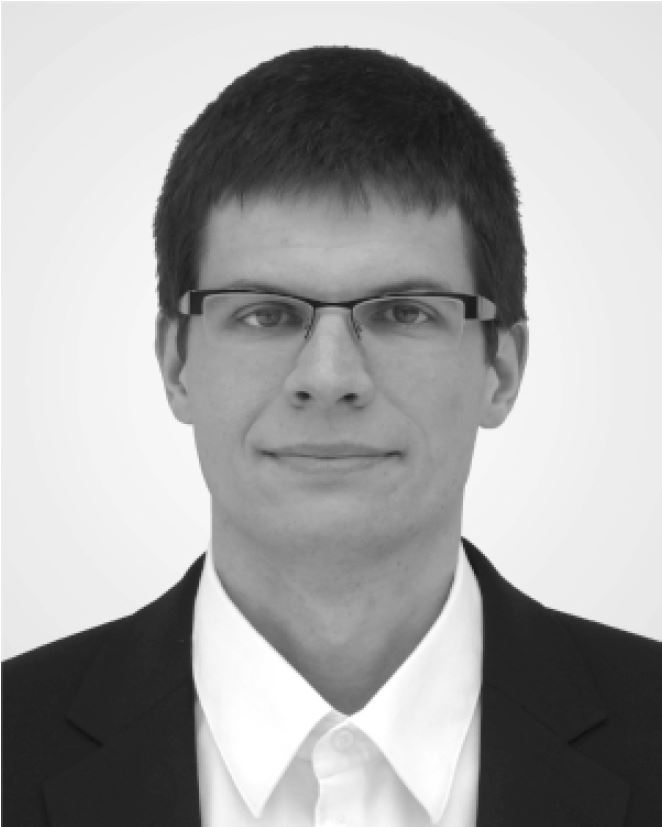}}]{Jonas Kornprobst}
	
	(S’17) received the B.Eng.
	degree in electrical engineering and information technology from the University of Applied Sciences
	Rosenheim, Rosenheim, Germany, in 2014, and the M.Sc. degree in electrical engineering and information technology from the Technical University of	Munich, Munich, Germany, in 2016. Since 2016, he has been a Research Assistant with
	the Chair of High-Frequency Engineering, Department of Electrical and Computer Engineering, Technical University of Munich. His current research interests include numerical electromagnetics, in particular integral equation	methods, antenna measurement techniques, antenna and antenna array design, as well as microwave circuits.
\end{IEEEbiography}
\begin{IEEEbiography}[{\includegraphics[width=1in,height=1.25in,clip,keepaspectratio]{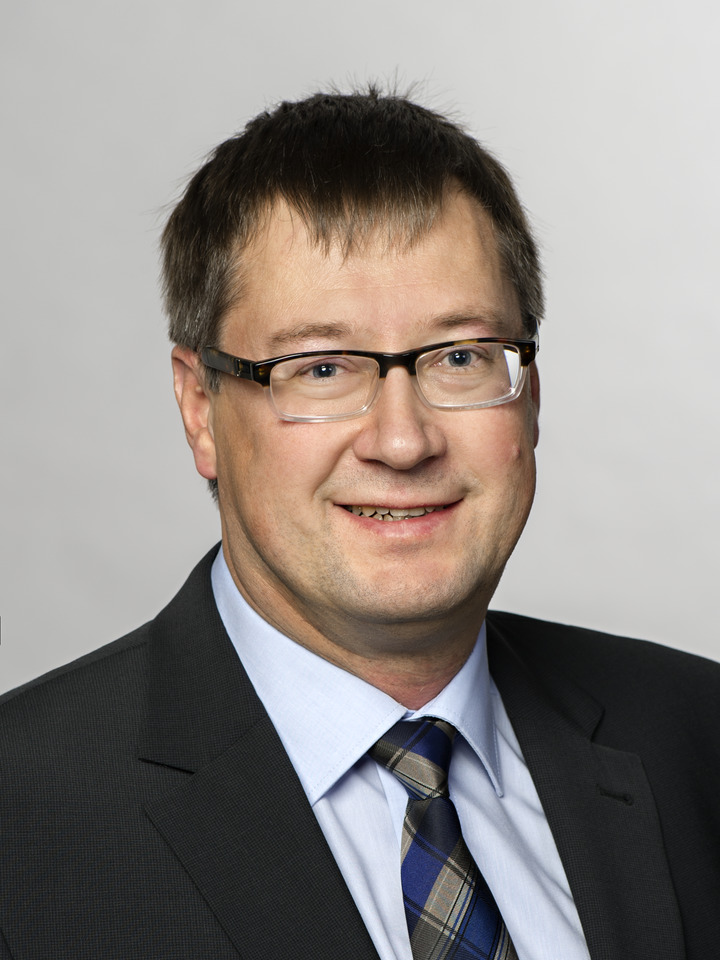}}]{Uwe Siart}
	
	(M'08) received the Dipl.-Ing. degree from the University of Erlangen-Nürnberg, Erlangen, Germany, in 1996 and the Dr.-Ing. degree from the Technical University of Munich, Munich, Germany, in 2005. He has been with the Chair of High-Frequency Engineering, Department of	Electrical and Computer Engineering, Technical University of Munich, since 1996. In 2005, he became a Senior Research Associate. His research interests are in the fields of signal processing and model-based parameter estimation for millimeter-wave radar signal	processing and high-frequency measurements. He is working on stochastic electromagnetic wave propagation, remote sensing the atmosphere, low-power radar sensors, microwave reflectometry and passive millimeter-wave components.		
\end{IEEEbiography}
\begin{IEEEbiography}[{\includegraphics[width=1in,height=1.25in,clip,keepaspectratio]{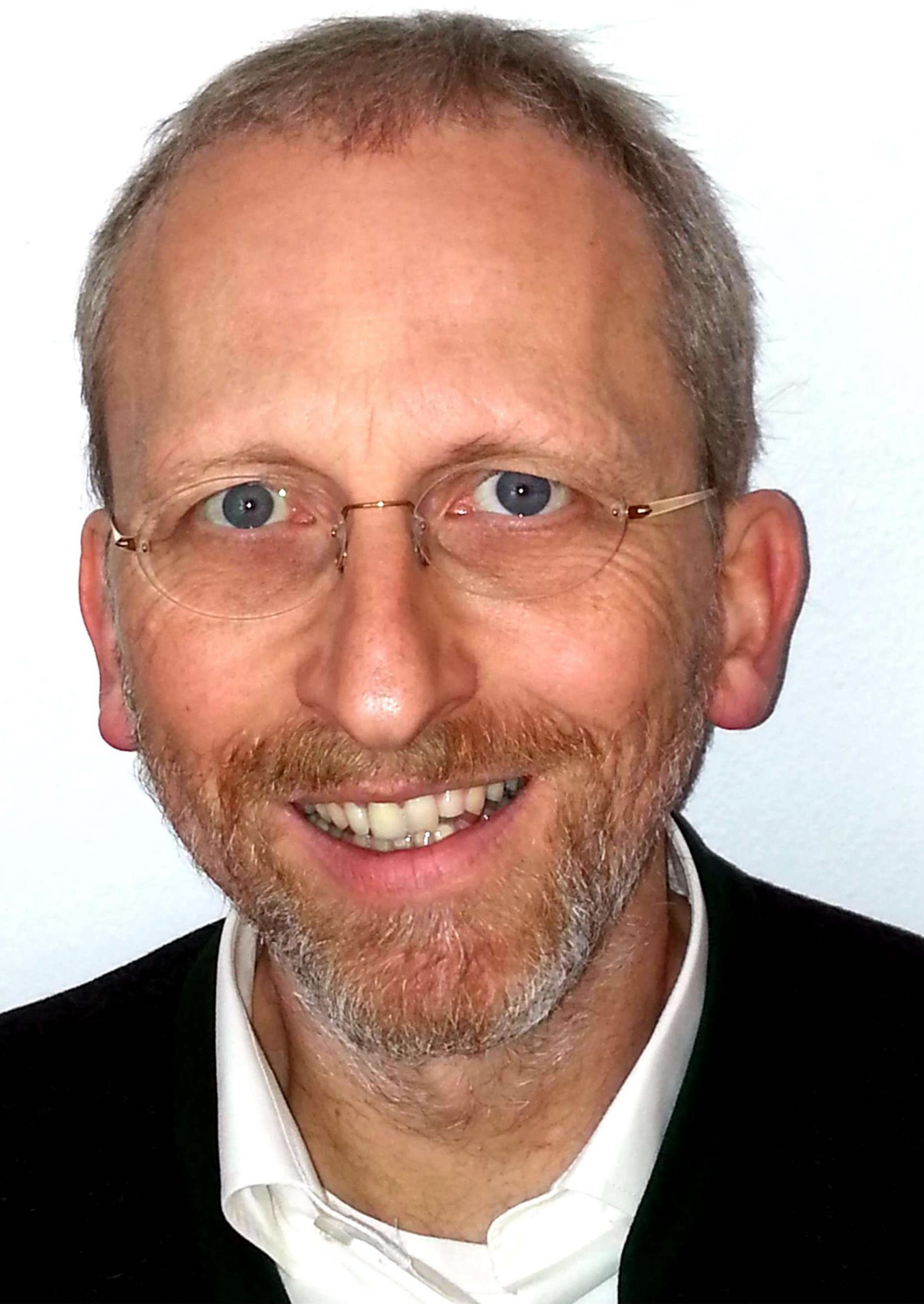}}]{Thomas F.~Eibert}
	
	(S'93$-$M'97$-$SM'09) received the Dipl.-Ing.\,\,(FH) degree from Fachhochschule N\"urnberg, Nuremberg, Germany, the Dipl.-Ing.~degree from Ruhr-Universit\"at Bochum, Bochum, Germany, and the Dr.-Ing.~degree from Bergische Universit\"at Wuppertal, Wuppertal, Germany, in 1989, 1992, and 1997, all in electrical engineering. He is currently a Full Professor of high-frequency engineering at the Technical University of Munich, Munich, Germany. His current research interests include numerical
electromagnetics, wave propagation, measurement and field transformation
techniques for antennas and scattering as well as all kinds of antenna and
microwave circuit technologies for sensors and communications.		
\end{IEEEbiography}
\end{document}